\documentclass[a4paper, fleqn, usenatbib, useAMS]{tex/mnras}

\usepackage{newtxtext, newtxmath}
\usepackage[T1]{fontenc}
\usepackage{ae, aecompl}

\usepackage{siunitx}

\usepackage{comment}

\usepackage{graphicx}
\graphicspath{{figs/}}
\DeclareGraphicsExtensions{.pdf, .ps, .eps, .jpg}
\usepackage[multidot]{grffile}

\usepackage{xspace}
\usepackage[usenames, dvipsnames]{xcolor}

\usepackage[normalem]{ulem}

\newcommand{\etal}{et\thinspace al.\thinspace}
\newcommand{\hii}{H\thinspace\textsc{ii}\xspace}
\newcommand{\starlight}{\textsc{starlight}\xspace}
\newcommand{\dobby}{\textsc{dobby}\xspace}

\newcommand{\Ha}{\ifmmode {\mathrm{H}\alpha} \else H$\alpha$\fi\xspace}
\newcommand{\Hb}{\ifmmode {\mathrm{H}\beta} \else H$\beta$\fi\xspace}
\newcommand{\Hg}{\ifmmode {\mathrm{H}\gamma} \else H$\gamma$\fi\xspace}

\newcommand{\oiii}{\ifmmode [\text{O}\,\textsc{iii}] \else [O~{\scshape iii}]\fi\xspace}
\newcommand{\Oiii}{\ifmmode [\text{O}\,\textsc{iii}]\lambda 5007 \else [O~{\scshape iii}]$\lambda 5007$\fi\xspace}
\newcommand{\nii}{\ifmmode [\text{N}\,\textsc{ii}] \else [N~{\scshape ii}]\fi\xspace}
\newcommand{\Nii}{\ifmmode [\text{N}\,\textsc{ii}]\lambda 6584 \else [N~{\scshape ii}]$\lambda 6584$\fi\xspace}
\newcommand{\Sii}{\ifmmode [\text{S}\,\textsc{ii}]\lambda 6716 \else [S~{\scshape ii}]$\lambda 6716$\fi\xspace}

\newcommand{\WHa}{\ifmmode W_{\mathrm{H}\alpha} \else $W_{\mathrm{H}\alpha}$\fi\xspace}

\title[The \Ha luminosity of galaxies with MaNGA and MUSE]
      {Less than the sum of its parts: the dust-corrected \Ha luminosity
      of star-forming galaxies explored at different spatial resolutions with MaNGA and MUSE}

\author[N.\ Vale Asari \etal]
       {N.\ Vale Asari,$^{1, 2}$\thanks{email: natalia@astro.ufsc.br}\thanks{Royal Society--Newton Advanced Fellowship}
        V.\ Wild,$^{2}$
        A.~L. de Amorim,$^{1}$
        A.\ Werle,$^{1, 3, 4}$
        Y.\ Zheng,$^{2}$
        \and
        R.\ Kennicutt,$^{5,6}$
        B.~D.\ Johnson,$^{7}$
        M.\ Galametz,$^{8}$
        E.~W.\ Pellegrini,$^{9}$
        R.~S.\ Klessen,$^{9,10}$
        \and
        S.\ Reissl,$^{9}$
        S.~C.~O.\ Glover,$^{9}$
        D.\ Rahner$^{9}$
        \and
        \\
        $^{1}$Departamento de F\'{\i}sica--CFM, Universidade Federal de Santa Catarina, C.P.\ 476, 88040-900, Florian\'opolis, SC, Brazil \\
        $^{2}$School of Physics and Astronomy, University of St Andrews, North Haugh, St Andrews KY16 9SS, UK\\
        $^{3}$Instituto de Astronomia, Geof\'{\i}sica e Ci\^{e}ncias Atmosf\'{e}ricas, Universidade de S\~{a}o Paulo, R. do Mat\~{a}o 1226, 05508-090 S\~{a}o Paulo, Brazil \\
        $^{4}$INAF - Osservatorio Astronomico di Padova, Vicolo dell'Osservatorio 5, 35122 Padova, Italy \\
        $^{5}$Steward Observatory, University of Arizona, 933 N Cherry Avenue, Tucson, AZ  85721-0065, United States\\
        $^{6}$Department of Physics \& Astronomy, Texas A\&M University, College Station, TX  77843-4242, United States\\
        $^{7}$Department of Astronomy, Harvard University, Cambridge, MA, United States\\
        $^{8}$AIM, CEA, CNRS, Universit\'{e} Paris-Saclay, Universit\'{e} Paris Diderot, Sorbonne Paris Cit\'{e}, F-91191 Gif-sur-Yvette, France\\
        $^{9}$Universit{\"a}t Heidelberg, Zentrum f{\"u}r Astronomie, Institut f{\"u}r Theoretische Astrophysik, Albert-Ueberle-Str. 2, 69120 Heidelberg, Germany\\
        $^{10}$Universit{\"a}t Heidelberg, Interdisziplin{\"a}res Zentrum f{\"u}r Wissenschaftliches Rechnen, Im Neuenheimer Feld 205,  69120 Heidelberg, Germany\\
      }

\date{Accepted \dots. Received \today; in original form \dots}

\pubyear{2019}

\begin{document}

\maketitle

\label{firstpage}
\pagerange{\pageref{firstpage}--\pageref{lastpage}}


\begin{abstract}
The \Ha and \Hb emission line luminosities measured in a single integrated spectrum are affected in non-trivial ways by point-to-point variations in dust attenuation in a galaxy.
This work investigates the impact of this variation when estimating global \Ha luminosities corrected for the presence of dust by a global Balmer decrement.
Analytical arguments show that the dust-corrected \Ha luminosity is always underestimated when using the global \Ha/\Hb flux ratio to correct for dust attenuation.
We measure this effect on 156 face-on star-forming galaxies from the Mapping Nearby Galaxies at APO (MaNGA) survey. At 1--2\,kpc spatial resolution, the effect is small but systematic, with the integrated dust-corrected \Ha luminosity underestimated by $2$--$4$ per cent (and typically not more than by $10$ per cent), and depends on the specific star formation rate of the galaxy. Given the spatial resolution of MaNGA, these are lower limits for the effect.
From Multi Unit Spectroscopic Explorer (MUSE) observations of NGC\,628 with a resolution of 36~pc we find the discrepancy between the globally and the point-by-point dust-corrected \Ha luminosity to be $14 \pm 1$ per cent, which may still underestimate the true effect.
We use toy models and simulations to show that the true difference depends strongly on the spatial variance of the \Ha/\Hb flux ratio, and on the slope of the relation between \Ha luminosity and dust attenuation within a galaxy.
Larger samples of higher spatial resolution observations are required to quantify the dependence of this effect as a function of galaxy properties.
\end{abstract}



\section{Introduction}
\label{sec:Introduction}

Although dust constitutes a tiny fraction of the interstellar medium, it may reprocess 30--99 per cent of the total stellar light in a normal disk galaxy (see the review by \citealp*{Galliano.Galametz.Jones.2018a} and references therein). Thus dust attenuation is often regarded as a nuisance, hindering our ability to measure the dust-free light which is key to understanding the star formation and chemical evolution of a galaxy.

The availability of well-calibrated integrated spectra of galaxies has encouraged the widespread use of simple recipes to account for dust attenuation. To correct the \Ha luminosity for dust (which may then be used as a proxy for the star formation rate in the last $\lesssim 10$~Myr), a single effective attenuation\footnote{We use the term dust attenuation in this work to account for the combined effect of dust geometry, scatter and absorption within a galaxy.} is typically inferred from a measured Balmer emission line flux ratio such as $f(\Ha)/f(\Hb)$, referred to as \Ha/\Hb or the Balmer decrement hereafter, alongside an assumed attenuation law. This is economical, since \Ha and \Hb can usually be measured in the same optical spectrum, which eschews flux cross-calibration problems. 

However, the chief limitation of this simple dust-correction recipe lies in ignoring point-by-point variations, since an integrated observation combines all the emitted-minus-absorbed and scattered light into a single spectrum. Point-by-point variations have been known for decades: in a seminal paper, \citet{Trumpler.1930b} found that dust is intermingled with the gas in the interstellar medium, and that total absorption varies towards various Milky Way line-of-sights. Measurements of the attenuation in the $V$ band ($A_V$) within a galaxy range from $0.37$--$2.2$\,mag in the core of NGC\,5253 \citep{Calzetti.etal.1997a} to 1--20\,mag for luminous and ultraluminous infrared galaxies \citep[LIRGS and ULIRGS, ][]{PiquerasLopez.etal.2013a}.

Point-by-point variations of $A_V$ in a galaxy imply that the globally-corrected \Ha is biased. Intuition suggests that the most dust-obscured regions contribute little to the integrated light, so the global Balmer decrement will be weighted towards the lowest values of \Ha/\Hb within a galaxy, and the dustiest regions will always be under-corrected. As this effect is not included in calibrations for star-formation rate derived from spectral synthesis models \citep[e.g.][]{Kennicutt.Evans.2012a}, this could result in a corresponding underestimation of star-formation rates from nebular emission lines. Similarly, metallicities derived from widely spaced emission lines will be affected. 

As we will show below, the amplitude of the effect depends on the non-trivial averaging of luminosity and Balmer decrement, with the added complication of potentially varying attenuation law slopes. Can observations help us quantify how close the intrinsic \Ha luminosity is to the dust-corrected \Ha from global spectra? \citet{Bassett.etal.2017c} have answered that question for a sample of 4 highly star-forming galaxies at $z = 0.05$--$0.15$ using \Ha maps from the Hubble Space Telescope and Paschen $\alpha$ from IFS data from the OH-Suppressing Infrared Integral field Spectrograph \citep{Larkin.etal.2006a} on the Keck telescope, with $\sim 1$~kpc resolution. They find a difference between locally-corrected and globally-corrected \Ha of $-5$ to $28$ per cent, with positive values indicating underestimated intrinsic \Ha from the globally-corrected \Ha. The largest effect is seen in the galaxy with the largest internal variations in $A_V$, which could either be intrinsic to the galaxy or, as they argue, due to its higher spatial resolution as it is also the lowest-redshift galaxy in their sample. Spatial resolution does seem to play a role in the size of the measured effect: \citet{PiquerasLopez.etal.2013a} found that after artificially increasing the spaxel scale for their LIRG and ULIRG sample from 40~pc to 0.2~kpc the measured median $A_V$ in their galaxies would be $\sim 0.8$ mag smaller.

In the following, we derive the amplitude of the effect as a function of the distribution of observed \Ha, Balmer decrement and attenuation law slope. We show that for a fixed dust attenuation law the global dust-corrected \Ha is always smaller than the point-by-point dust-corrected \Ha. We assess the accuracy of dust-corrected integrated \Ha luminosities using two sets of optical integral field spectroscopic (IFS) data, both of which cover the \Ha and \Hb line regions. The first set is a sample of 156 star-forming galaxies from the Mapping Nearby Galaxies at APO \citep[MaNGA;][]{Bundy.etal.2015a,Drory.etal.2015a} survey. The most recent public data release of MaNGA has over 4800 galaxies with optical IFS data at 1--2~kpc resolution, which has allowed us to cull an unprecedentedly large high-quality sample of star-forming galaxies encompassing a range of specific star formation rates (sSFR). The second set are datacubes of the galaxy NGC\,628 obtained with the Multi Unit Spectroscopic Explorer \citep[MUSE;][]{Bacon.etal.2010a} with a spatial resolution of $\sim 40$~pc. Those datacubes for a single galaxy allow us to investigate the impact of spatial resolution on the dust-corrected \Ha. 

For readers in a hurry, the key plot of this work is Fig.~\ref{fig:LHa-corr}.
This paper is organised as follows.
Section~\ref{sec:model} lays out a simple analytical framework to quantify the amplitude of the effect.
Sections~\ref{sec:manga} and \ref{sec:muse} describe our data analysis for MaNGA and MUSE IFS cubes, and our MaNGA sample selection.
Section~\ref{sec:LILG} compares the global to the point-by-point dust-corrected \Ha for
MaNGA star-forming galaxies and for NGC\,628 observed with MUSE. 
In Section~\ref{sec:discussion} we discuss our results in light of simple toy models and galaxy simulations, and investigate the impact of the diffuse ionized gas (DIG) and varying dust laws.
We summarise our findings in Section~\ref{sec:summary}.

\section{Analytical approach}
\label{sec:model}

Our aim is to quantify the effect of observing a variety of regions in a galaxy with distinct physical properties in a single beam. Given a distribution of regions with different properties (intrinsic $\Ha$ luminosity and dust attenuation), the single-beam integrated spectrum contains a non-trivial averaging of $\Ha$ luminosity and $\Ha/\Hb$ flux ratio, which affects the total dust-attenuated \Ha flux. Implicit in this approach is the fact that the emission lines can be powered by different ionization sources; here we are simply interested in the averaging effect.

Let us first define the parameters associated with the global integrated spectrum (thus the subscript G below). The intrinsic $\Ha$ luminosity is obtained from
\begin{align}
L_\mathrm{G} = L_\mathrm{G}^\mathrm{obs} \, e^{+ \tau_\mathrm{G}},
\end{align}
where $L_\mathrm{G}^\mathrm{obs}$ is the observed $\Ha$ luminosity and $\tau_\mathrm{G}$ is the optical depth at the wavelength of $\Ha$. $\tau_\mathrm{G}$ is calculated from the Balmer decrement, $L_\alpha^\mathrm{obs}/L_\beta^\mathrm{obs}$, as
\begin{align}
  \tau_\mathrm{G} = - \dfrac{1}{1 - q} \,\ln \dfrac{
  L_\mathrm{G,\alpha}^\mathrm{obs}/ L_\mathrm{G,\beta}^\mathrm{obs} }{ L_\alpha/L_\beta },
\end{align}
where $L_\alpha/L_\beta$ is the intrinsic $\Ha/\Hb$ luminosity ratio (which in this work we assume to be 2.87 for the Case B, i.e.\ assuming a nebula that is an optically thick to Lyman photons, at 10\,000\,K and low density; \citealp{Osterbrock.Ferland.2006a}), and $q$ is the value at \Hb of an attenuation curve normalised at \Ha, which depends on the shape of the attenuation curve:
\begin{align}
\label{eq:q}
L_\mathrm{G,\beta} = L_\mathrm{G,\beta}^\mathrm{obs} \, e^{+ \tau_\mathrm{G} \, q}.
\end{align}
Note that we use $\alpha$ and $\beta$ subscripts to disambiguate the meaning of $L_\mathrm{G}$; where missing, we assume it refers to \Ha.

Let us now consider a galaxy made up of $N$ individual regions, each with its own observed and intrinsic \Ha luminosities ($l^\mathrm{obs}$ and $l$). The total intrinsic \Ha luminosity of those individual regions is
\begin{align}
L_\mathrm{IFS} = \sum_{j=1}^N \, l_{j} =  \sum_j l^\mathrm{obs}_j e^{+\tau_{j}},
\end{align}
where $\tau_{j}$ is the optical depth at $\Ha$ for the $j$th region given by
\begin{align}
\tau_j = - \dfrac{1}{1 - q_j} \,\ln \dfrac{
  l_\mathrm{j,\alpha}^\mathrm{obs} / l_\mathrm{j,\beta}^\mathrm{obs} }{ L_\alpha/L_\beta }
\end{align}
and $q_j$ is the value at \Hb of the attenuation curve, normalised at \Ha, for that region of the galaxy. 

We want to compare the intrinsic \Ha luminosity corrected point-by-point, $L_\mathrm{IFS}$, to the one measured from the integrated spectrum, $L_\mathrm{G}$.  Noting that $L_\mathrm{G,\alpha}^\mathrm{obs} = \sum_j l_\mathrm{j,\alpha}^\mathrm{obs}$, and defining
\begin{align}
\label{eq:wj}
w_j \equiv \dfrac{l_\mathrm{j,\alpha}^\mathrm{obs}}{\sum_j l_\mathrm{j,\alpha}^\mathrm{obs}},
\end{align}
we can write
\begin{align}
\dfrac{L_\mathrm{IFS}}{L_\mathrm{G}} = \dfrac{\sum_j w_j e^{+\tau_j}}{e^{+\tau_\mathrm{G}}}.
\end{align}

After some mathematical manipulation, one can show that
\begin{align}
\label{eq:LIFS_LG}
\dfrac{L_\mathrm{IFS}}{L_\mathrm{G}}
= \dfrac{\sum_j w_j e^{+\tau_j}}{\left[\sum_j w_j e^{+\tau_j (1-q_j)}\right]^{1/(1-q)}}.
\end{align}
In the case where the attenuation curves do not vary ($q_j = q$), both the numerator and the denominator are weighted power means (also known as H\"older power means) of the type $M_m = \left( \sum_j w_j x^m \right)^{1/m}$, for which the inequality $M_m > M_n$ holds when $m > n$. With $m=1$, $n=1-q$ and $x=e^{\tau_j}$ we see that $L_\mathrm{IFS}>L_\mathrm{G}$ when $q>0$. Given the definition of $q$ (equation~\ref{eq:q}), this will hold for any attenuation law\footnote{Even for pure scattering, which could lead to $0<q<1$ \citep*[e.g.][]{Witt.Thronson.Capuano.1992a}.}. For completeness, we note that $q \sim 1.4$ for standard dust extinction or attenuation laws such as \citet*[hereafter CCM]{Cardelli.Clayton.Mathis.1989a} with $R_V = 3.1$ and \citet*{Calzetti.Kinney.Storchi-Bergmann.1994a}. While the more general case of varying $q_j$ values cannot be proven analytically to lead to $L_\mathrm{IFS}>L_\mathrm{G}$ in all cases, a simple numerical test shows that the inequality usually holds for reasonable assumed values of $q_j$.  

\defcitealias{Cardelli.Clayton.Mathis.1989a}{CCM}

Equation~(\ref{eq:LIFS_LG}) shows how $L_\mathrm{IFS}/L_\mathrm{G}$  is inextricably linked to the slope(s) of the attenuation laws in the galaxy, as well as the relative distributions of gas, dust and stars. The down-weighting of regions with large $\tau_j$ will have an even stronger impact if these regions also have steeper attenuation curves than average. Any underestimation in the global $q$, whether estimated or measured, will also lead to a larger effect. We explore further the impact of gas, dust and star geometry on the amplitude of the $L_\mathrm{IFS}/L_\mathrm{G}$ ratio in Section \ref{sec:spatial-distribution} below. 

In the following, we demonstrate the effect exists in observations by measuring ${L_\mathrm{IFS}}/{L_\mathrm{G}}$ for MaNGA galaxies. MaNGA itself has limited spatial resolution, and each one of its fibres is in itself an integrated spectrum. We therefore check the impact of spatial resolution with MUSE observations of NGC\,628, and use toy models and simulations to estimate the true magnitude of the effect.

\section{Data processing and sample selection of MaNGA galaxies}
\label{sec:manga}

MaNGA \citep{Bundy.etal.2015a} is one of the spectroscopic programs of the fourth phase of the  Sloan Digital Sky Survey (SDSS, \citealp{York.etal.2000a}; SDSS-IV, \citealp{Blanton.etal.2017a}). Upon its completion, the project will have collected datacubes for about 10\,000 galaxies up to redshift $\sim0.03$ with spatial resolution of 1--2 kpc. MaNGA observations are made with hexagonal bundles of 19 to 127 2-arcsec-diameter fibres, and the spectra cover the 3600--10400 \AA\ wavelength range with spectral resolution $R\sim2000$. In this work, we use 4824 MaNGA galaxy datacubes available from the 15th SDSS data release \citep{Aguado.etal.2019a}.

Redshifts are available from the NASA-Sloan Atlas \citep[NSA, ][]{Blanton.etal.2011a} catalogue, and taken from the MaNGA \texttt{drpall} table version 2.4.3.  We assume a flat $\Lambda$CDM cosmology, with $\Omega_0 = 0.3$ and $H_0 = 70\,\text{km}\,\text{s}^{-1}\,\text{Mpc}^{-1}$. The luminosities and the physical scale of the pixels are based on the luminosity and angular diameter distances calculated from the redshift.

Our pre-processing steps, spectral continua fit and emission line measurements are the same as detailed in \citet{ValeAsari.etal.2019a}. In the following we summarise the most crucial steps.

\subsection{Preprocessing and spectral continuum fitting}
\label{sec:data}

Our preprocessing starts by masking out whole spaxels (due to low fibre coverage or foreground stars) and wavelengths with bad pixels. 
Since the FWHM of the PSF is about 2.5 arcsec, cubes are binned to a sampling of $1 \times 1$ arcsec per spaxel (i.e.\ a $2 \times 2$ binning).
Spectra are corrected for Galactic extinction assuming a  \citetalias{Cardelli.Clayton.Mathis.1989a} law with $R_V = 3.1$, with $E(B - V)$ from \cite*{Schlegel.Finkbeiner.Davis.1998a} dust maps.    
They are then shifted to the rest frame and resampled to 3600--10400~\AA\ with $\Delta\lambda = 1\,\text{\AA}$. 
MaNGA spectra are converted from vacuum to air wavelengths to match the spectral synthesis models.

The preprocessed MaNGA spectra are then fitted with the stellar population synthesis code \starlight \citep{CidFernandes.etal.2005a}. \starlight 
models the observed spectrum pixel-by-pixel as a sum of stellar populations. 
We mask emission lines, bad pixels and sky features, and assume a single dust screen following the \citetalias{Cardelli.Clayton.Mathis.1989a} law with $R_V = 3.1$. The fits are made on the 3600--8900~\AA\ spectral range to avoid residuals from bright OH sky lines in the red part of the spectra. 

We use 96 stellar populations of constant star-formation rate (SFR) in log-age bins between 1 Myr and 14 Gyr and metallicities $Z = 0.0001$--$0.05$, calculated from the 2016 models by \citet{Bruzual.Charlot.2003a}\footnote{\url{http://www.bruzual.org/~gbruzual/cb07/Updated_version_2016/}}.
We have chosen models with a \citet{Chabrier.2003a} initial mass function, `Padova (1994)' \citep{Alongi.etal.1993a,Bressan.etal.1993a,Fagotto.etal.1994a,Fagotto.etal.1994b,Girardi.etal.1996a} evolutionary tracks and `MILES' stellar library \citep{SanchezBlazquez.etal.2006a, FalconBarroso.etal.2011a}. In fact, the MILES library is used below 7351 \AA, and complemented by the STELIB library \citep{LeBorgne.etal.2003a} above that wavelength. Observed stars are further complemented by theoretical spectra from \citet{Martins.etal.2005a}, Tlusty \citep{Lanz.Hubeny.2003a, Lanz.Hubeny.2003b}, UVBlue \citep{RodriguezMerino.etal.2005a} and PoWR \citep{Sander.Hamann.Todt.2012a}. We have verified that the results of this paper are insensitive to the exact details of the stellar population models used.

\subsection{Balmer emission line measurements}\label{sec:Balmer}

We obtain the nebular residual spectra by subtracting the stellar continuum models from the observed spectra.  Our emission-line fitting (ELF) code \dobby\footnote{\dobby is a free ELF available as part of the {\scshape pycasso2} package at \url{https://bitbucket.org/streeto/pycasso2}.} fits a Gaussian profile to emission lines in the residual spectra.  We impose kinematic ties for lines emitted by the same ion, and we also require that the $\Ha/\Hb$ line flux ratio is $\geq 2.6$ and $\Nii/\nii\lambda 6548 = 3$. The code first fits $\nii\lambda\lambda 6548,6584$ and \Ha simultaneously as they may be blended, then fits \Hb using the systemic velocity and intrinsic velocity dispersion from \Ha. 

The total velocity dispersion is calculated by adding the intrinsic and the instrumental dispersions in quadrature. Equivalent widths are given by  $W = F / C$, where $F$ is the emission line flux and $C$ is the continuum flux density at the line centre measured in the \starlight synthetic spectra. We define the amplitude-to-noise ratio ($A/N$) for an emission line as the ratio between the Gaussian amplitude and $\sigma_N$, which is the rms in the residual continuum contiguous to the emission line. The signal-to-noise ratio is given by  $S/N = F / (\sigma_N \sqrt{6 \sigma_\lambda \Delta\lambda})$, where the flux uncertainty is based on \cite{Rola.Pelat.1994a}, with $\sigma_\lambda$ being the Gaussian dispersion in \AA, and $\Delta\lambda = 1$~\AA\ the spectral sampling. 

Our results rely on differences between emission line fluxes of the order of a few per cent, and therefore it is necessary to take extreme care about small systematic errors. By comparing the observed \Hb flux from the integrated spectrum with the sum of \Hb spaxel-by-spaxel for galaxies with good enough S/N in \Hb we were able to identify two important effects:

\begin{enumerate}

\item There is a well-known and ubiquitous trough in the residual spectra following stellar continuum fitting around \Hb (e.g.\ seen in SDSS, e.g.\ \citealp{CidFernandes.2006a}; \citealp*{Groves.Brinchmann.Walcher.2012a}, and CALIFA, \citealp{CidFernandes.etal.2014a}). We first modelled this trough with a straight line, which occasionally caused an underestimate in the \Hb fluxes. Fitting the pseudo-continuum with a Legendre polynomial of degree 16 as close to the emission line as possible proved more satisfactory.
  
\item We initially set a single instrumental dispersion of $70$~km/s at all wavelengths, the average dispersion for MaNGA cubes.  Due to the MaNGA instrumental dispersion being greater at \Hb than at \Ha, this caused an underestimation in the width, and therefore flux, of the \Hb line. The full dispersion spectra from the MaNGA cubes were required to obtain the correct measurement of \Hb flux.
  
\end{enumerate}

Both effects, taken together, caused \Hb to be underestimated by $\sim 3$ per cent on average, but larger variations of up to $\sim 10$ per cent were detected in individual spaxels. The measurements used throughout our analysis have been corrected for these two effects. Furthermore, we have ensured that measurements of \Hb fluxes with the kinematic ties switched on or off (the latter with two more degrees of freedom) are identical within uncertainties. The advantage of kinematic ties between \Hb and \Ha is that they help the code correctly measure \Hb fluxes in noisier spectra, thus increasing our sample of galaxies with useful \Hb measurements in all spaxels.

\subsection{Galaxy sample selection}
\label{sec:sample}

Our master sample is defined very similarly to the one by \citet{ValeAsari.etal.2019a}. For clarity, we repeat the criteria used and highlight the one difference to their selection.

\begin{enumerate}
\item By requiring at least one spectrum in the cube with $S/N \geq 3$ in the 5590--5680 \AA\ spectral region, and that galaxies have redshifts $> 0$ in the NSA catalogue, 4676 datacubes were fitted with both \starlight and \dobby. 

\item We then kept only those in the Primary+ and Secondary MaNGA
subsamples with data quality flag \texttt{DRP3QUAL} of 0 (see \citealp{Law.etal.2016a}, table 16, updated online at \url{https://www.sdss.org/dr13/algorithms/ bitmasks/#MANGA_DRP3QUAL}). 

\item Duplicate observations of the same galaxy were removed, keeping the cube with the largest number of fibres, and with the largest exposure time in the case where more than one cube was observed with the same fibre bundle.

\item We removed edge-on galaxies by selecting galaxies with axial ratio $b/a \geq 0.3$, measured by the NSA using elliptical apertures.

\item We then matched our galaxies to the Galaxy Zoo I catalogue \citep[GZ, ][]{Lintott.etal.2011a}, and removed those which have a probability of being in a merging system $\leq 0.4$ \citep{Darg.etal.2010a}, and also those with no GZ  classification.

\item We selected only galaxies with $> 80$ unmasked spaxels within one $R_{50}$, where $R_{50}$ is the $r$-band Petrosian 50 per cent light radius (as measured by the NSA). This is the only difference to the selection by \citet{ValeAsari.etal.2019a}, who used $R_{50}$ from the seventh data release of the SDSS. This excludes galaxies with a large number of masked spaxels due to foreground objects. 

\end{enumerate}

These cuts lead to a master sample of 3\,185 galaxies.
Finally, we select a sample of 156 star-forming galaxies with extremely high-quality observations from this master sample, with $A/N > 2$ in \Ha and \Hb in all $2 \times 2$ binned spaxels, and integrated \Nii/\Ha versus \Oiii/\Hb emission line ratios placing them below the \cite{Stasinska.etal.2006a} line that delineates the region of pure star-forming galaxies.
A caveat to bear in mind is that our sample of high-quality observations is biased against galaxies with low sSFR. Nevertheless, in order to correctly measure the impact on dust corrections of spatial averaging, which is extremly sensitive to noise, we have decided to favour galaxies observed with a high signal-to-noise ratio rather than adding faint galaxies.

\section{MUSE observations of NGC\,628}
\label{sec:muse}

In order to investigate the extent to which our study depends on the $\sim$~kpc MaNGA spatial resolution, we repeat our analysis for NGC\,628, a nearby Sc galaxy \citep{deVaucouleurs.etal.1991a} observed with MUSE \citep{Bacon.etal.2010a} at the Very Large Telescope. The telescope archival data is available via the ESO Phase 3 Data Release, reduced with the MUSE pipeline version \texttt{muse-1.4} or higher \citep{Weilbacher.etal.2012b, Weilbacher.etal.2014a,  Weilbacher.Streicher.Palsa.2016a}. We downloaded the 12 MUSE non-overlapping datacubes with the highest exposure times (ranging from 42 to 50 min), each one covering a $1' \times 1'$ field-of-view and $4750$--$9350$\AA\ spectral range with $R \sim 3000$. We realigned the datacubes to correct for offsets in the astrometry \citetext{K.\ Kreckel, priv. comm.}, and removed overlapping spaxels. We resampled the datacubes from $0.2$ to $0.8$-arcsec spaxels to match the typical seeing (\citealp{Kreckel.etal.2016a, Kreckel.etal.2018a}; the seeing ranges from $0.7$ to $1.1$ arcsec), leading to a spatial sampling of 36~pc. A total of 62\,175 spaxels with $S/N \geq 3$ around 5635 \AA\ were analysed.

The preprocessing steps and fits with \starlight and \dobby have been kept as similar as possible to the MaNGA pipeline described above. We use the recession velocity of $657$~km/s \citep{Lu.etal.1993a} to shift the spectra to the rest frame, and adopt a luminosity distance of 9.7~Mpc \citep{Dhungana.etal.2016a} and a Galactic selective extinction of $E(B - V) = 0.062$ from \citet{Schlafly.Finkbeiner.2011a}. \starlight fits are performed in the 4750--8900 \AA\ range. Emission lines are measured in the residual (observed minus \starlight model) spectra with \dobby assuming an instrumental resolution of $R = 2000$ at \Hb and $R = 4000$ at \Ha\footnote{The variation of the spectral resolution as a function of $\lambda$ can be found in the MUSE User Manual v.~7.3, \url{https://www.eso.org/sci/activities/vltsv/muse/ESO-261650_7_MUSE_User_Manual.pdf}.}.
Similarly to the MaNGA data, setting the correct spectral dispersion is crucial to avoid underestimating the \Hb line flux. Spectra at the edges of the 12 datacubes proved to be noisy, so to avoid biases we have removed all 2\,735 edge spaxels from our analysis, which would contribute to $\sim 3$ of the total \Ha and \Hb observed luminosities.

About 7 per cent of the remaining spaxels (4\,332 out of 59\,440) have $A/N \leq 2$ in either \Ha or \Hb, so we stack their spectra and remeasure the emission lines in this stacked spectrum. This low-$A/N$ spectrum contributes only $0.8$ and $0.6$ per cent to the total \Ha and \Hb observed luminosities. The low-$A/N$ stacked spectrum and all other high-$A/N$ spaxels are individually taken into account to calculate $L_\mathrm{IFS}$.

In Sec.~\ref{sec:muse-results} below we also show results for NGC\,628 with downgraded spatial sampling. Each of the 12 cubes has been individually downsampled, and spaxels at the edges are removed up to the 0.15~kpc-sampling cubes (where they amount to $\sim 5$ per cent of the total \Ha and \Hb observed luminosities).

\section{Results: The spatially dust-corrected \Ha luminosity}
\label{sec:LILG}

We can now tackle the question this work set out to answer: how accurate is the global dust-corrected \Ha in comparison to the point-by-point dust-corrected \Ha? For each object in our sample of 156 MaNGA star-forming galaxies and for NGC\,628 observed with MUSE we calculate the dust-corrected \Ha luminosity from the global integrated spectrum ($L_\mathrm{G}$) and also the sum of the dust-corrected \Ha luminosities spaxel-by-spaxel ($L_\mathrm{IFS}$). Uncertainties in luminosities are estimated from formal error propagation. From equation~(\ref{eq:LIFS_LG}) our prediction is that $L_\mathrm{IFS}/L_\mathrm{G} \geq 1$.

\begin{figure*}
  \centering
  \includegraphics[width=1\textwidth, trim=260 50 160 0, clip]{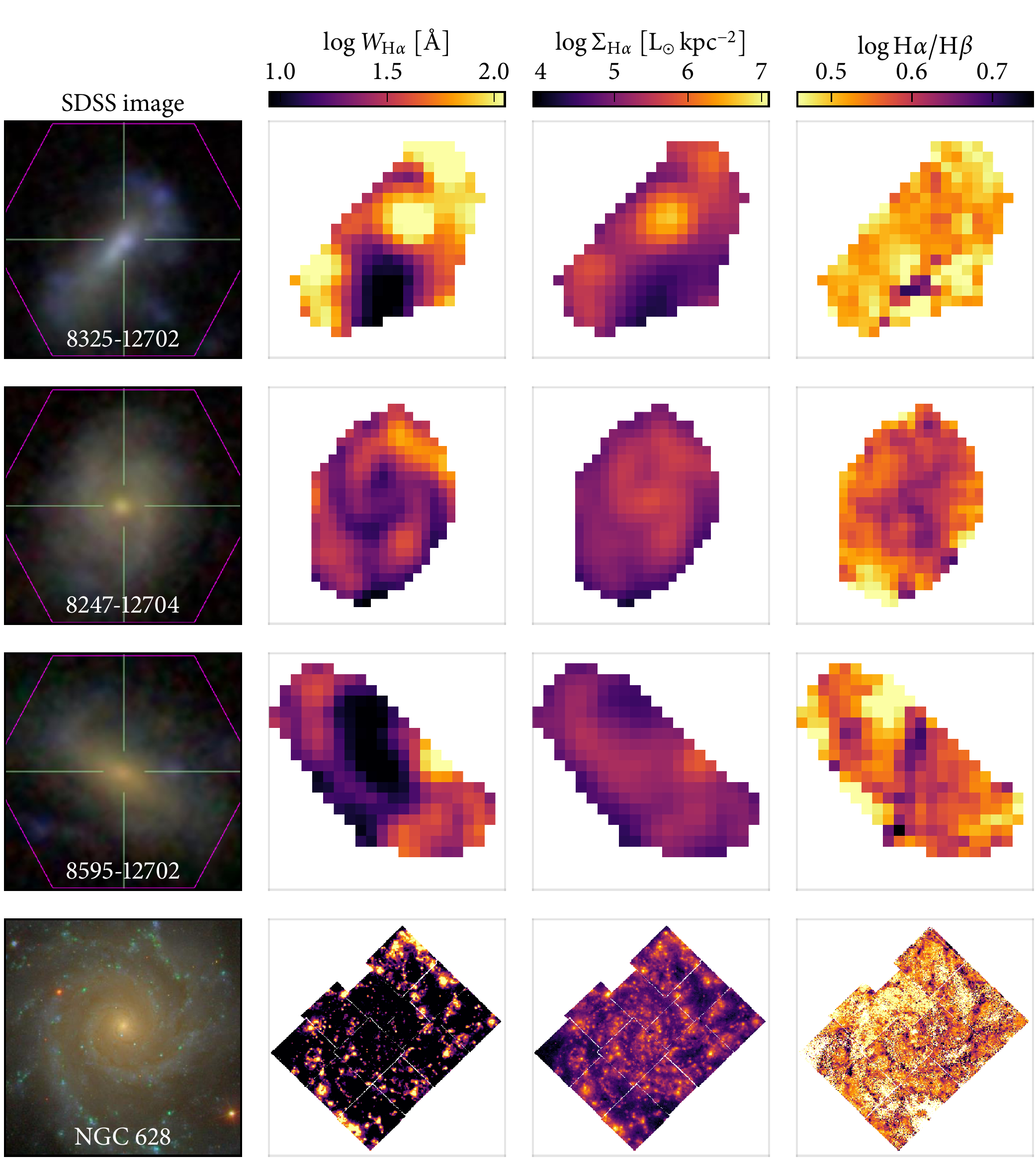}
  \caption{Top three rows: three example galaxies from our MaNGA star-forming sample, with different emission and dust distributions. Bottom row: NGC\,628 observed with 12 MUSE pointings and a spatial sampling of $36$~pc.
    The left column shows the 3-colour SDSS image, and for the MaNGA observations we show the \texttt{plate-ifudsgn} identification and the fibre bundle outlined in pink. The other columns show maps of the equivalent width of \Ha, the observed \Ha surface density, and the Balmer decrement (\Ha/\Hb).
    For NGC\,628, spaxels with $A/N \leq 2$, which have been stacked onto a single low-$A/N$ spectrum, are masked out (in white) on the \Ha/\Hb map.
    From top to bottom, the physical linear size of the grey squares delimiting the maps are 19, 27, 20 and 11~kpc. For all images and maps throughout this paper, North is at the top and East to the left.}
\label{fig:examples}
\end{figure*}

Fig.~\ref{fig:examples} shows images and emission line measurements for three example galaxies in our MaNGA star-forming sample and for NGC\,628 observed with MUSE, with maps of \WHa, the observed \Ha luminosity surface density, and \Ha/\Hb. In the following we calculate $L_\mathrm{IFS}$ by correcting the observed \Ha surface density using the spaxel-by-spaxel \Ha/\Hb map. $L_G$ is calculated by correcting the global-spectrum $\Ha$ luminosity with the global-spectrum \Ha/\Hb for each galaxy. In both cases we assume a \citetalias{Cardelli.Clayton.Mathis.1989a} dust law with $R_V = 3.1$ and set $\tau=0$ when\footnote{We note that the floor in \Ha/\Hb applied in Section \ref{sec:Balmer} and the requirement for non-negative $\tau$ values when dust correcting our luminosities presents the potential for one-sided errors, which could result in asymmetric scatter. We therefore verified that $L_\mathrm{IFS}/L_\mathrm{G}$ does not change noticeably for any galaxy if we remove these restrictions. Similarly, we have tested assigning different values for the intrinsic \Ha/\Hb ratio (ranging from 2.80 to 3.10) to spectra with different \Nii/\Ha ratios (a proxy for the nebular metallicity), we find $L_\mathrm{IFS}/L_\mathrm{G}$ increases on average by 0.005, so this does not significantly affect our results. } \Ha/\Hb $< 2.87$. 

\begin{figure*}
  \centering
  \includegraphics[width=.8\textwidth, trim=0 205 340 20, clip]{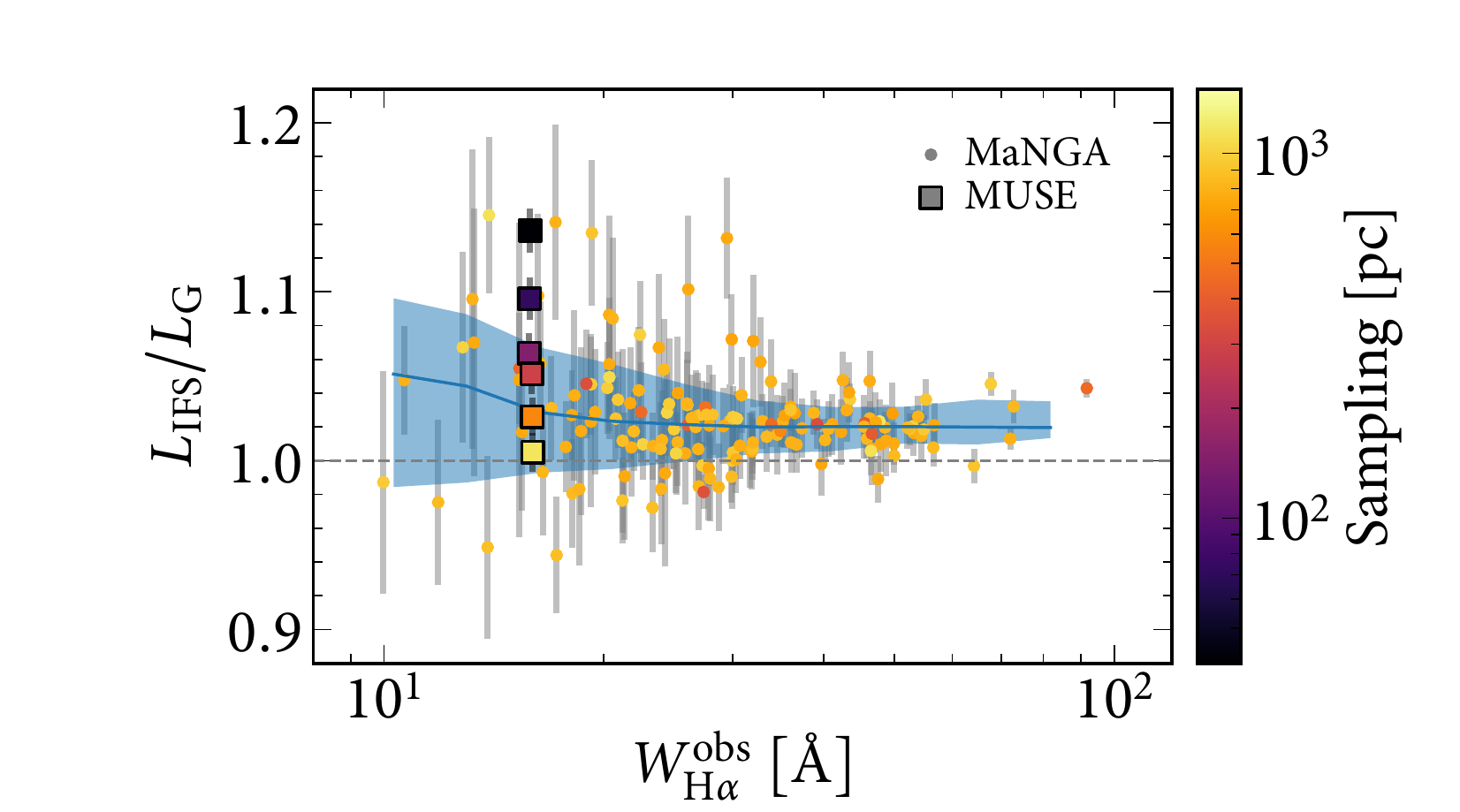} 
  \caption{The ratio of the \Ha luminosity corrected spaxel-by-spaxel for dust attenuation to the value corrected by the global \Ha/\Hb, $L_\mathrm{IFS}/L_\mathrm{G}$, as a function of the global observed \WHa, which is correlated with the sSFR and anticorrelated with the dominance of DIG emission in a galaxy.
    The 156 MaNGA star-forming galaxies are shown as dots, and NCG\,628 observed with MUSE as squares, with the different colours indicating the linear physical scale of the spaxels. The MUSE observations range from a 36~pc sampling up to 1.1~kpc, doubling for each consecutive point.
    Error bars show the uncertainty in the ratio from formal error propagation.
    The blue line and shaded area show the rolling median and 16 and 84 percentiles for the MaNGA objects. The dashed horizontal line marks $L_\mathrm{IFS}/L_\mathrm{G} = 1$. We note that $L_\mathrm{IFS}/L_\mathrm{G} \geq 1$ for 153 MaNGA galaxies within their one-sigma uncertainty.  Larger discrepancies between the dust-corrected \Ha from global spectra as compared to the more accurate dust-corrected \Ha measured spaxel-by-spaxel are found for galaxies with smaller \WHa: the average $L_\mathrm{IFS}/L_\mathrm{G}$ for MaNGA objects varies from $1.037 \pm 0.008$ to $1.021 \pm 0.002$ for galaxies in the lowest and highest $\WHa^\mathrm{obs}$ quartile in our sample.
  }
\label{fig:LHa-corr}
\end{figure*}

\subsection{Results for MaNGA star-forming galaxies}
\label{sec:manga-results}

Fig.~\ref{fig:LHa-corr} shows $L_\mathrm{IFS}/L_\mathrm{G}$ as a function of the global observed \Ha equivalent width, $\WHa^\mathrm{obs}$, with galaxies colour-coded by the linear physical spaxel size. The blue line and shaded area indicate the rolling median and 16 and 84 percentiles calculated at every 0.1 dex with a kernel of 0.5 dex. The squares refer to the MUSE observations of NGC\,628, which we will discuss in Section \ref{sec:muse-results} below.
Our high-quality sample of star-forming galaxies requires marginally good point-by-point detections of \Ha and \Hb, which yields $L_\mathrm{IFS}/L_\mathrm{G}$ ratios with acceptable uncertainty values. One must again bear in mind that this procedure selects against low $\WHa^\mathrm{obs}$ ($\lesssim 20$ \AA), thus low sSFR objects.

We immediately see that $L_\mathrm{IFS}/L_\mathrm{G}$ is indeed $\geq 1$ for most (153 out of 156) galaxies in our MaNGA sample within their one-sigma uncertainty. Additionally, at low $\WHa^\mathrm{obs}$ the distribution is lopsided, and it is galaxies with small $\WHa^\mathrm{obs}$ that are more likely to have a larger measured $L_\mathrm{IFS}/L_\mathrm{G}$ of up to 15 per cent. 

In order to check for trends in $L_\mathrm{IFS}/L_\mathrm{G}$, we have divided our sample into $\WHa^\mathrm{obs}$ quartiles, and compared statistical properties among those bins.
The average $L_\mathrm{IFS}/L_\mathrm{G}$ values for the bins range from $1.037 \pm 0.008$ to $1.021 \pm 0.002$ from the first to the last quartile, where the errors are from propagation of errors on the individual points. We warn that the uncertainties quoted are not hypothesis-free, and refer the reader to Appendix~\ref{app:stats} for a more robust statistical comparison among bins.
To investigate whether the scatter changes with $\WHa^\mathrm{obs}$ we calculate the sample standard deviation propagating the errors on the individual data points. The scatter ranges from $0.048\pm 0.008$ to $0.013\pm 0.002$ with increasing $\WHa^\mathrm{obs}$. Finally, we use an Anderson-Darling test \citep{Press.etal.2007a} to calculate the probability $p$ that the low and high \WHa bins are drawn from the same distribution, obtaining $p \leq 0.001$. The aforementioned Appendix~\ref{app:stats} also shows trends of $L_\mathrm{IFS}/L_\mathrm{G}$ with other physical and observational parameters.

Our results imply that galaxies with low $\WHa^\mathrm{obs}$, or equivalently low sSFR, are more likely to have high values of $L_\mathrm{IFS}/L_\mathrm{G}$. In Sections \ref{sec:spatial-distribution} and \ref{sec:toy-models} we investigate which properties of the luminosity-dust spatial relationship might cause high values of $L_\mathrm{IFS}/L_\mathrm{G}$ in certain galaxies. 
Low values of $\WHa^\mathrm{obs}$ suggest a larger contribution to the total line emission from the DIG \citep{Lacerda.etal.2018a, ValeAsari.etal.2019a}. In Section \ref{sec:DIG} below, we investigate whether the increased number of galaxies with high $L_\mathrm{IFS}/L_\mathrm{G}$ at small $\WHa^\mathrm{obs}$ might be due to a higher proportion of emission from the DIG.

We additionally investigated trends with parameters other than $\WHa^\mathrm{obs}$, which are not shown here. There is a slight increase in $L_\mathrm{IFS}/L_\mathrm{G}$ at stellar surface mass densities below $10^{7.5}$M$_\odot$/kpc$^2$, which also tend to have low $\WHa^\mathrm{obs}$, although galaxies with the highest $L_\mathrm{IFS}/L_\mathrm{G}$ have a wide range of mass densities. The galaxies with the highest $L_\mathrm{IFS}/L_\mathrm{G}$ values were observed with the largest IFU sizes, however, there were no clear trends with inclination ($b/a$), the physical scale of spaxels (which range from 0.3 to 1.2~kpc), number of spaxels, NSA concentration index or the maximum Petrosian radius included in the observations.  

\subsection{Results for MUSE observations of NGC\,628}
\label{sec:muse-results}

It may seem reassuring that $L_\mathrm{IFS}/L_\mathrm{G}$ for MaNGA galaxies is typically 2--4 per cent, and rarely greater than 10 per cent; however, we emphasise this is a lower limit. While MaNGA allows us to select a large sample of star-forming galaxies, its spatial resolution is of the order of 1--2~kpc, much larger than individual \hii regions. Thus each spaxel must suffer the same statistical averaging as our global integrated spectrum, and our measurements of the spaxel-by-spaxel corrected $L_\text{IFS}$ are limited by the MaNGA spatial resolution.

Overplotted in Fig.~\ref{fig:LHa-corr} are the results for MUSE observations of NGC\,628 (squares). The black square shows that $L_\mathrm{IFS}/L_\mathrm{G} = 1.14 \pm 0.01$ for the spatial sampling of 36~pc. We have also downgraded the cubes to 0.072, 0.15, 0.29, 0.58 and 1.2~kpc samplings (i.e.\ doubling the spaxel linear size each time), the latter two covering the typical range of samplings of our MaNGA star-forming galaxies. The other squares show that $L_\mathrm{IFS}/L_\mathrm{G} =
1.10;
1.06;
1.05;
1.03;
1.00 \pm 0.01$ for the degraded MUSE data.
We see no convergence for the high-resolution data, which may suggest that even higher resolution data may be needed to correctly measure $L_\mathrm{IFS}/L_\mathrm{G}$.
These results show that the difference between the globally dust-corrected and the locally dust-corrected \Ha luminosities would likely be significantly higher if our MaNGA star-forming sample were observed with a $10$--$100$~pc resolution.

\section{Discussion}
\label{sec:discussion}

Our empirical results show that the \Ha luminosity corrected for the global \Ha/\Hb is indeed underestimated when compared to the \Ha luminosity that has been corrected spaxel-by-spaxel. 
In this section, we investigate the physical reasons behind the variation in $L_\mathrm{IFS}/L_\mathrm{G}$, as well its true amplitude, with the help of toy models and simulations.

\subsection{The spatial distribution of dust attenuation}
\label{sec:spatial-distribution}

\begin{figure*}
  \centering
  \includegraphics[width=.8\textwidth, trim=250 160 270 10, clip]{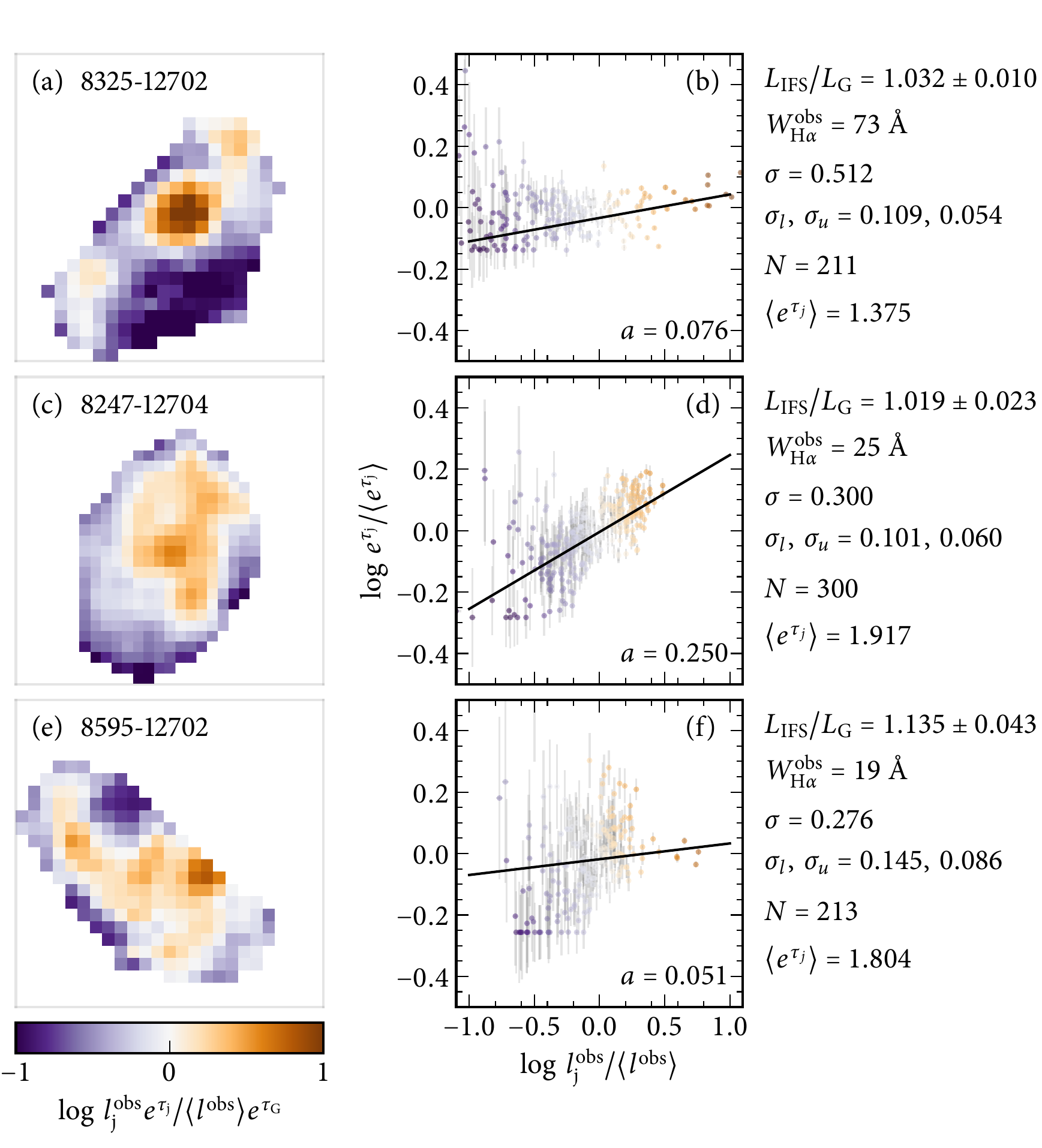}
  \caption{Left panels: For the same three example galaxies in Fig.~\ref{fig:examples}, we show maps of $l^\mathrm{obs}_j e^{+\tau_j} / \langle l^\mathrm{obs}_j \rangle e^{+\tau_\mathrm{G}} $, which indicates how much or how little a region contributes to $L_\mathrm{IFS}/L_\mathrm{G}$ (see equation~\ref{eq:LIFS_LG}). Right panels: mean-normalised observed \Ha flux versus the mean-normalised dust attenuation in each spaxel, colour-coded as the maps on the left.  We fit a regression line to the points taking into account the uncertainties, and show the slope $a$ in the bottom right.  To the right of the plots, we show $L_\mathrm{IFS}/L_\mathrm{G}$, the global \WHa, the scatter $\sigma$ in the $l^\mathrm{obs}_j/\langle l^\mathrm{obs}_j\rangle$ distribution, the scatter $\sigma_l$ and $\sigma_u$ in $e^{+\tau_j} / \langle e^{+\tau_j} \rangle$ for lower and upper values of $l^\mathrm{obs}_j/\langle l^\mathrm{obs}_j\rangle$, the number $N$ of spaxels, and the mean attenuation $\langle e^{+\tau_j} \rangle$.}
\label{fig:toy1}
\end{figure*}

What causes the variation in $L_\mathrm{IFS}/L_\mathrm{G}$ between galaxies? We can begin to address this question by studying the spatially-resolved MaNGA maps in detail.

The left panels of Fig.~\ref{fig:toy1} show, for the same MaNGA galaxies from Fig.~\ref{fig:examples}, maps of $l^\mathrm{obs}_j e^{+\tau_j} / \langle l^\mathrm{obs}_j\rangle e^{+\tau_\mathrm{G}}$, which indicates how much a region contributes to $L_\mathrm{IFS}/L_\mathrm{G}$ (see equation~\ref{eq:wj} and \ref{eq:LIFS_LG}). Relevant global quantities are given on the right-hand side, including $L_\mathrm{IFS}/L_\mathrm{G}$, \WHa and the average optical depth $\langle e^{+\tau_j} \rangle$. The remaining quantities will be described in Section~\ref{sec:toy-models}.

The scatter plots on the right show, for each spaxel, the mean-normalised observed \Ha flux ($l^\mathrm{obs}_j/\langle l^\mathrm{obs}_j \rangle$)\footnote{The choice of using $l^\mathrm{obs}_j / \langle l^\mathrm{obs}_j \rangle \equiv N w_j$ (see equation~\ref{eq:wj}) instead of $l_j / \langle l_j \rangle$ is twofold. First, the former is an observational quantity independent of the choice of a dust attenuation law.
    Second, writing $l^\mathrm{obs}_j / \langle l^\mathrm{obs}_j \rangle$ allows one to prove from equation~\ref{eq:LIFS_LG} that $L_\mathrm{IFS}/L_\mathrm{G} \ge 1$ for the specific case where the dust law does not vary across a galaxy. Throughout this discussion we have thus favoured to quote $l^\mathrm{obs}_j/\langle l^\mathrm{obs}_j \rangle$ values when possible.
}
versus the mean-normalised dust attenuation ($e^{+\tau_j} / \langle e^{+\tau_j} \rangle$). Values with large $l^\mathrm{obs}_j/\langle l^\mathrm{obs}_j\rangle$ and small $e^{+\tau_j} / \langle e^{+\tau_j} \rangle$ will dominate the integrated flux from the galaxy. The regression line takes into account uncertainties in the ordinate and abscissa, and points are colour-coded as in the map on the left. The slope and scatter of this relation are key to understanding individual $L_\mathrm{IFS}/L_\mathrm{G}$ values for each galaxy, and we will explore them using toy models below.

These example galaxies have been picked to highlight the different distributions of dust and nebular emission in galaxies, which we describe in turn:

\begin{itemize}

\item 8325-12702 is an irregular, highly star forming system, with low level, smoothly distributed dust attenuation and a large $\WHa^\mathrm{obs}$ for our sample (the median $\WHa^\mathrm{obs}$ value is 28\AA). The predominant contribution to $L_\mathrm{IFS}/L_\mathrm{G}$ comes from the central star-forming region of the galaxy, where the dust attenuation is representative of the whole galaxy. The dusty clump to the lower right (south-southwest) of the galaxy is in a region of low level star formation, and the off-centre star-forming clumps in lower dust attenuation regions are not sufficiently bright to outshine the central region. Therefore $L_\mathrm{IFS}/L_\mathrm{G} - 1$ is small ($\sim3$ per cent).

\item 8247-12704 is a spiral galaxy with a small central bulge and spiral arms visible in both the \WHa and \Ha maps, and dust distributed co-spatially with the nebular emission. The dominant contribution to $L_\mathrm{IFS}/L_\mathrm{G}$ comes from the star-forming central region and arms of the galaxy, where the dust is smoothly distributed. Therefore $L_\mathrm{IFS}/L_\mathrm{G} - 1$ is small ($\sim2$ per cent).

\item 8595-12702 is a barred spiral, with a global $\WHa^\mathrm{obs}$ that is on the low side for our sample. The prominent off-centre star-forming region dominates the $L_\mathrm{IFS}/L_\mathrm{G}$. However, the dust content of this region is considerably lower than in the star-forming arms of the galaxy. This leads to an underestimated global \Ha/\Hb and large $L_\mathrm{IFS}/L_\mathrm{G} - 1$ ($\sim14$ per cent). Galaxies such as this one would have a larger aperture bias in their global \Ha luminosities when observed with fiber spectroscopy, such as the Sloan Digital Sky Survey \citep{York.etal.2000a}.

\end{itemize}

While 8595-12702 and 8247-12704 have very similar dust contents and sSFR, they have very different 
$L_\mathrm{IFS}/L_\mathrm{G}$ values. Without the off-centre dust-poor star-forming region 8595-12702 might be expected to have a similarly low $L_\mathrm{IFS}/L_\mathrm{G}$. Likewise, if 8595-12702 had a much more active star forming central region, as in 8325-12702, the off-centre region would have less impact on the total luminosity and therefore the $L_\mathrm{IFS}/L_\mathrm{G}$ would be lower.

\subsection{Toy models}
\label{sec:toy-models}

\begin{figure*}
  \centering
  \includegraphics[width=1.0\textwidth, trim=100 180 100 0]{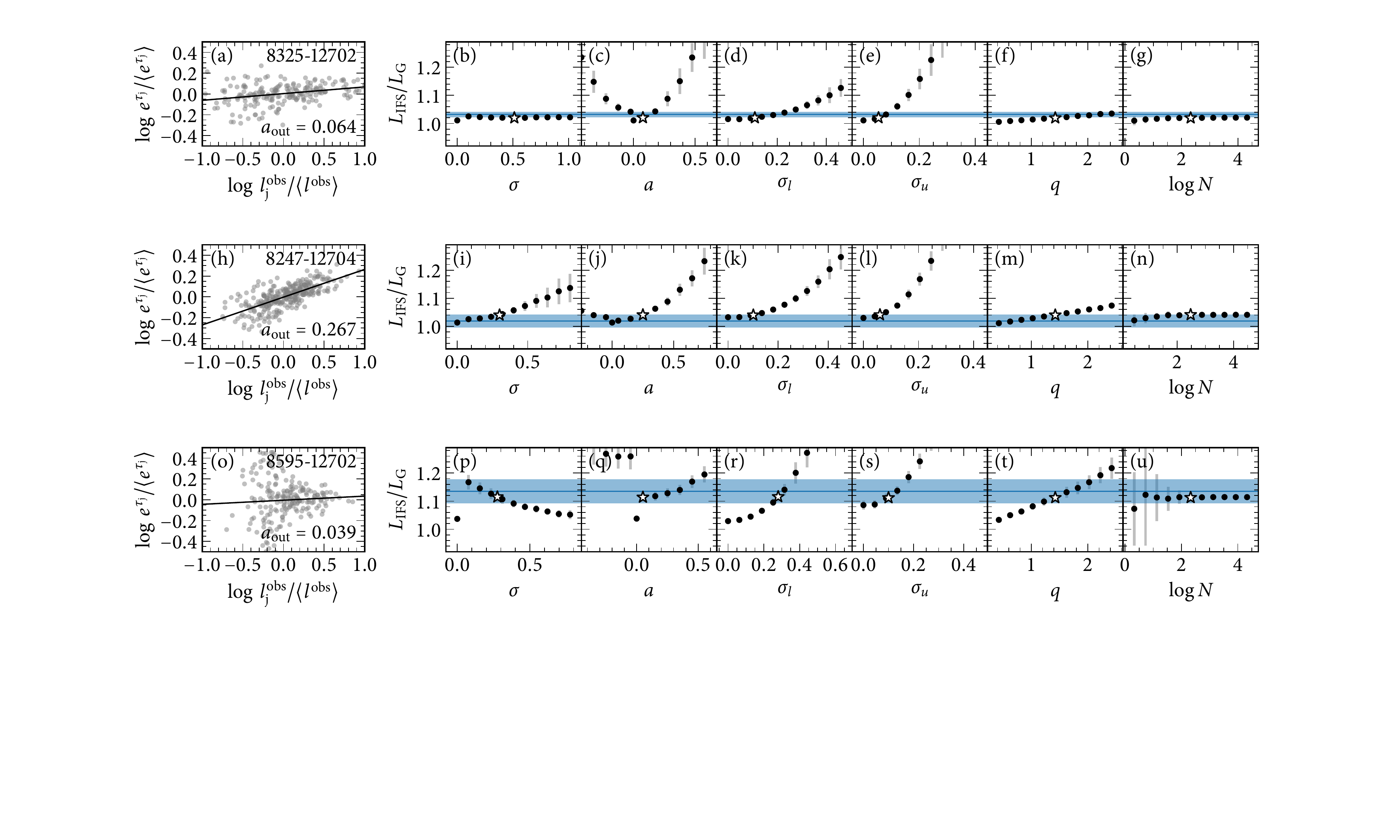}
  \caption{Toy models tailored for the three example MaNGA galaxies.  Panels on the left show one realisation of the $l^\mathrm{obs}_j/\langle l^\mathrm{obs}_j\rangle$ versus $e^{+\tau_j} / \langle e^{+\tau_j} \rangle$ relation, and the line regression to the points gives the output slope $a_\mathrm{out}$. Panels to the right show the variation of $L_\mathrm{IFS}/L_\mathrm{G}$ as a function of the input parameters: the scatter $\sigma$ in $\log l^\mathrm{obs}_j/\langle l^\mathrm{obs}_j\rangle$, the input slope $a$ in the log-linear relation between luminosity and attenuation, the scatter $\sigma_l$ and $\sigma_u$ for the lower and upper values of the \Ha flux, the steepness $q$ of the dust law, and the number of spaxels $N$. Fiducial values, marked with open stars, are kept constant while we vary one input parameter at a time.  Horizontal lines and shaded areas in blue indicate the measured $L_\mathrm{IFS}/L_\mathrm{G}$ value and uncertainty for the MaNGA galaxies. }
\label{fig:toy-models}
\end{figure*}

In order to understand which parameters have a larger influence upon $L_\mathrm{IFS}/L_\mathrm{G}$, we have created toy models guided by the MaNGA data, and in particular the scatter plots presented in the right panels of Figure \ref{fig:toy1}. We first draw values for $\log l^\mathrm{obs}_j/\langle l^\mathrm{obs}_j\rangle$ from a Gaussian distribution with dispersion $\sigma$ for $N$ galaxy spaxels. We take a log-linear relation with slope $a$ to map the $l^\mathrm{obs}_j/\langle l^\mathrm{obs}_j\rangle$ onto $e^{+\tau_j} / \langle e^{+\tau_j} \rangle$ values. We then add scatter to $e^{+\tau_j} / \langle e^{+\tau_j} \rangle$.  The scatter is typically larger at small values of the normalised \Ha flux, so our models include two values for the scatter, $\sigma_l$ and $\sigma_u$, below and above $l^\mathrm{obs}_j/\langle l^\mathrm{obs}_j\rangle = 1$. For the purposes of these models it is not important whether the increased scatter at low luminosities is intrinsic or caused by measurement errors. We additionally vary $q$, the slope of the attenuation curve, noting that $q<1$ implies an inverted attenuation curve, in which \Hb is less attenuated than \Ha, which might occur in some regions due to scattering, but is not expected when averaging over large spatial areas. 

The left panels of Fig.~\ref{fig:toy-models} show three different realisations of the toy model using the empirical values for $\sigma$, $a$, $\sigma_l$, $\sigma_u$ and $N$ for the three example galaxies from Fig.~\ref{fig:toy1}, assuming $q=1.42$ from the \citetalias{Cardelli.Clayton.Mathis.1989a} dust law with $R_V = 3.1$. The other panels show the effect of varying one parameter at a time, leaving the others at the fiducial empirical values (open stars), to see how this affects the final result. We create 100 realisations for each set of toy models with fixed parameters, so our points are the mean and the error bars the rms of our Monte Carlo simulations. The horizontal line and shaded area show the measured $L_\mathrm{IFS}/L_\mathrm{G}$ value and uncertainty for each example galaxy. The value $a_{\rm out}$ given in the left panel is remeasured from the toy models, so it is a little different from the input value due to the noise and difference between the shape of the real and modelled distributions. Because of the complexity of the distribution for galaxy 8595-12702, which is not captured by our simple toy model, we set $\sigma_l = 0.280$ and $\sigma_u = 0.100$ to make the  $l^\mathrm{obs}_j/\langle l^\mathrm{obs}_j\rangle$ versus $e^{+\tau_j} / \langle e^{+\tau_j} \rangle$ scatter plot more similar to the observed one.
Note how the fiducial parameters are able to reproduce the empirical $L_\mathrm{IFS}/L_\mathrm{G}$ within the uncertainties.

The figure shows that changing either the slope, $a$, or the scatter at large relative luminosities, $\sigma_u$, has a large effect in all cases regardless of the other parameters. A steep positive or negative correlation between luminosity and dust attenuation will result in a mismatch between the global versus spatially-resolved \Ha/\Hb, and therefore a large disparity between $L_\mathrm{IFS}$ and $L_\mathrm{G}$. Similarly, a very large scatter between luminosity and dust attenuation in the few high luminosity spaxels will invariably lead to a large discordance. When there is little correlation between dust attenuation and luminosity ($a$ is close to zero), and little scatter in that relation at high luminosities ($\sigma_u$ is close to zero), the other quantities have little effect and the difference remains small (see top row, galaxy 8325-12702). Panels (i) and (p) show that the dispersion in the luminosity, $\sigma$, can either correlate or anticorrelate with  $L_\mathrm{IFS}/L_\mathrm{G}$. As predicted in Section \ref{sec:model} $L_\mathrm{IFS}/L_\mathrm{G}$ decreases as $q$ decreases, but remains greater than 1 at all times. 

The seemingly null correlation between $L_\mathrm{IFS}/L_\mathrm{G}$ and the number of spaxels $N$ may look puzzling, given that the effect is much greater in our MUSE observation with high resolution than in both the downgraded MUSE cubes and in the MaNGA galaxies. This however is due to a limitation in our toy models: when decreasing the number of spaxels, we do not add spaxels together, but simply generate $N$ new values of $l^\mathrm{obs}_j/\langle l^\mathrm{obs}_j\rangle$ and $e^{+\tau_j} / \langle e^{+\tau_j} \rangle$ using the recipe above. This highlights that the increased effect at high spatial resolution is a real physical effect, and not simply due to there being more data points to reduce the impact of random fluctuations. Section~\ref{sec:simulations} below uses a simulated galaxy in order to investigate the real effect of binning spaxels together.

Ultimately, these toy models highlight the major difficulty with this problem: the models contain non-trivial dependencies between their parameters, making it difficult to predict $L_\mathrm{IFS}/L_\mathrm{G}$ for a particular galaxy.
For the irregular highly star-forming galaxy 8325-12702, the only parameters that could increase $L_\mathrm{IFS}/L_\mathrm{G}$ were the slope of the luminosity-attenuation relation and the scatter in the relation at high luminosities.
On the other hand, for the regular spiral 8247-12704, changing any of the parameters could result in a larger $L_\mathrm{IFS}/L_\mathrm{G}$. 

\subsection{The impact of the DIG}
\label{sec:DIG}

Fig.~\ref{fig:LHa-corr} indicates that the galaxies with the largest values of $L_\mathrm{IFS}/L_\mathrm{G}$ all have low $\WHa^\mathrm{obs}$. We now turn to the question of whether $L_\mathrm{IFS}/L_\mathrm{G}$ is expected to be higher for star-forming galaxies with the lowest values of $\WHa^\text{obs}$ due to an increased contribution of DIG emission. DIG can arise from gas beyond the star-forming \hii regions, and therefore is expected to show less spatial correlation with either the nebular emission or the dust attenuation (thus impacting $a$ and $\sigma_l$). It may also exhibit a different attenuation law ($q$), and lead to an increased dispersion in the relation between luminosity and dust ($\sigma$).

\begin{figure}
  \centering
  \includegraphics[width=1.0\columnwidth, trim=25 55 230 20, clip]{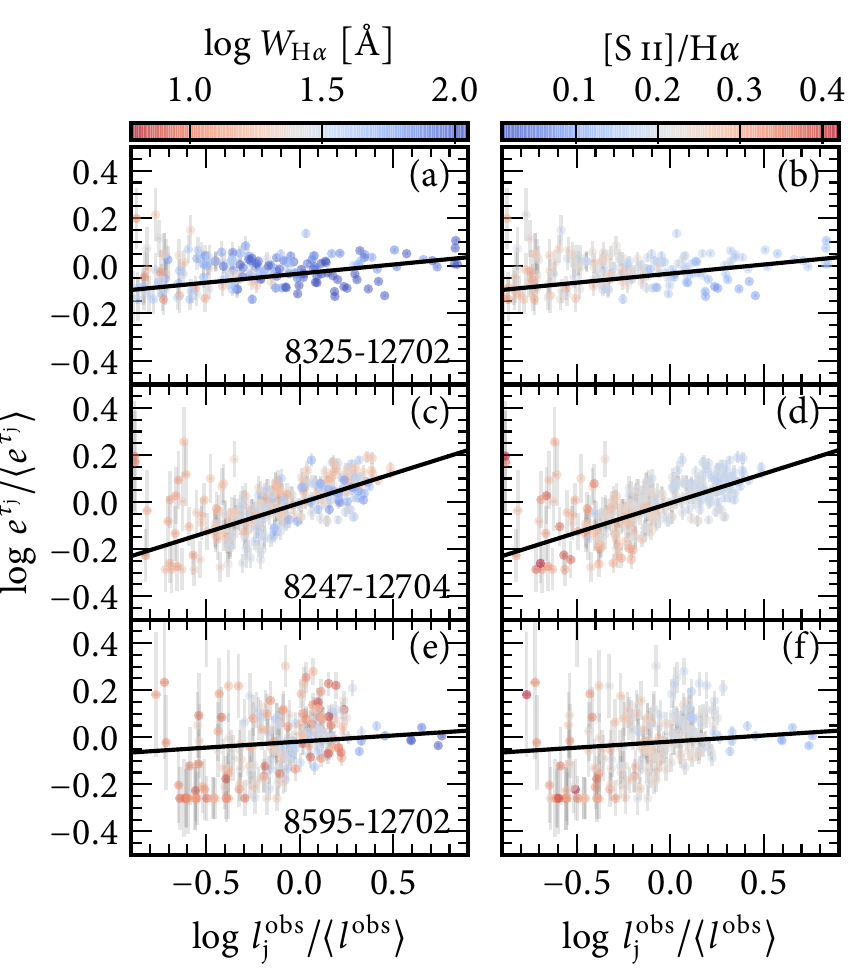}
  \caption{Plots show $l^\mathrm{obs}_j/\langle l^\mathrm{obs}_j\rangle$ versus $e^{+\tau_j} / \langle e^{+\tau_j} \rangle$ colour-coded by \WHa (left) and \Sii/\Ha (right) in each spaxel for the three example MaNGA galaxies.  Both parameters are related with the importance of DIG in a spaxel: lower values of \WHa and higher values of \Sii/\Ha are indicative of a larger contribution from DIG to the emission line luminosities, thus points in red mean a greater DIG contribution according to each parameter.}
\label{fig:WHa-S2Ha}
\end{figure}

Fig.~\ref{fig:WHa-S2Ha} shows again the $l^\mathrm{obs}_j/\langle l^\mathrm{obs}_j\rangle$ versus $e^{+\tau_j} / \langle e^{+\tau_j} \rangle$ for all spaxels in our three example MaNGA galaxies, but now colour-coded by \WHa (left) or \Sii/\Ha (right). Lower values of \WHa and higher values of \Sii/\Ha are indicative of a larger contribution from DIG to the emission line luminosities \citep[e.g.][]{Reynolds.1985a, Blanc.etal.2009a, Lacerda.etal.2018a, Poetrodjojo.etal.2019a, ValeAsari.etal.2019a},
although both can be affected by secondary effects such as stellar content, gas density and metallicity, making it important to use multiple tracers. The two galaxies with the lowest global equivalent widths, 8247-12704 and 8595-12702, have a higher fraction of spaxels with larger DIG contributions than 8325-12702, with these spaxels dominating at low relative \Ha luminosities.
Where there is likely significant contribution from the DIG, the scatter at fixed relative \Ha luminosity is noticeably larger. While some of this increase in scatter is clearly due to the larger error bars at low line luminosities, there is clearly additional intrinsic scatter in 8595-12702. In this case, the increased scatter extends above $l^\mathrm{obs}_j/\langle l^\mathrm{obs}_j\rangle=1$ and may well play a role in the large $L_\mathrm{IFS}/L_\mathrm{G}$.

Even though our sample of MaNGA galaxies is biased against low global values of \WHa, many of their spaxels still show a non-negligible contribution from DIG emission according to several criteria used to detect DIG. \citet{Lacerda.etal.2018a} and \citet{ValeAsari.etal.2019a} define DIG-dominated spaxels as $\WHa < 14$~\AA\  for CALIFA \citep{Sanchez.etal.2012a,Sanchez.etal.2016a} and MaNGA observations. \citet{Blanc.etal.2009a} consider that the DIG contributes to $> 50$ per cent of the \Ha emission in spaxels where $\Sii/\Ha > 0.15$ for galaxy NGC~5194 (see their fig.~9) observed with the Visible Integral field Replicable Unit Spectrograph Prototype (VIRUS-P). Finally, \citet{Zhang.etal.2017c} define a cut at $\Sigma_\Ha > 10^{5.4} \mathrm{\,L_\odot\, kpc^{-2}}$ to select \hii region-dominated spaxels in MaNGA, so the complementary criterion would define zones with DIG contribution. Fig.~\ref{fig:examples} and \ref{fig:WHa-S2Ha} show for our three example galaxies, which span a large range of global \WHa values, that indeed all of them have at least some spaxels dominated by DIG emission according to these criteria.

\begin{figure}
  \centering
  \includegraphics[width=\columnwidth, trim=80 72 650 50, clip]{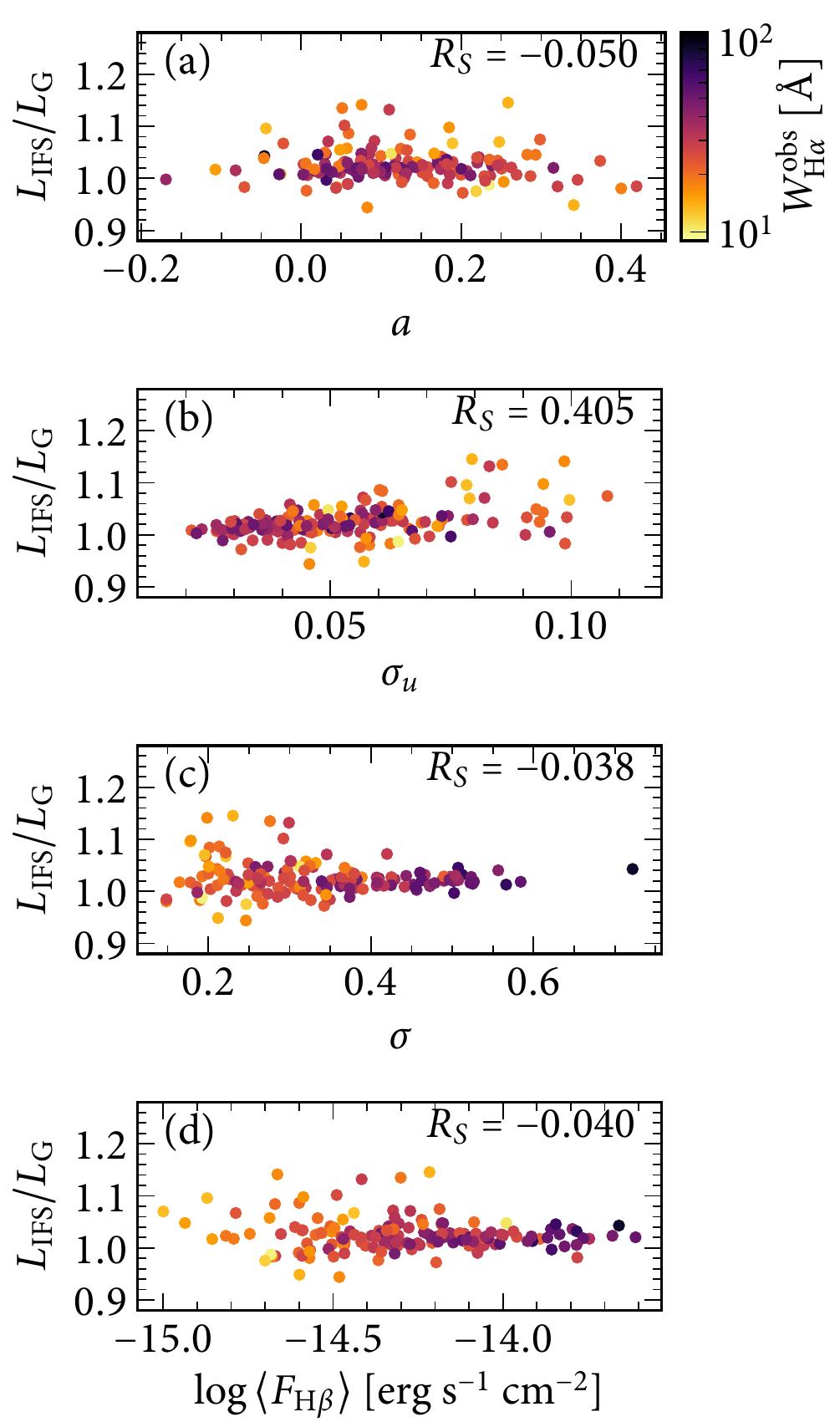}
  \caption{$L_\mathrm{IFS}/L_\mathrm{G}$ for the MaNGA star-forming galaxies as a function of the most important parameters used in the toy models: (a) the slope in the $l^\mathrm{obs}_j/\langle l^\mathrm{obs}_j\rangle$ versus $e^{+\tau_j} / \langle e^{+\tau_j} \rangle$ relation; (b) the scatter $\sigma_u$ around this relation for high-luminosity spaxels; (c) the scatter in the values of $l^\mathrm{obs}_j/\langle l^\mathrm{obs}_j\rangle$; and (d) the average of \Hb line flux measured in the spaxels. Points are colour-coded by the global \WHa, and on the top right-hand side of each panel we show the values of the Spearman rank correlation coefficient.}
\label{fig:DIG}
\end{figure}

In order to investigate patterns within the whole sample, Fig.~\ref{fig:DIG} shows $L_\mathrm{IFS}/L_\mathrm{G}$ as a function of the most important parameters used in the toy models  for all galaxies in the MaNGA sample, colour-coded by $\WHa^\mathrm{obs}$. The slope of the relation, $a$, appears to be largely uncorrelated with either $\WHa^\mathrm{obs}$ or $L_\mathrm{IFS}/L_\mathrm{G}$ (panel a), although the largest positive values do tend to be found in galaxies with lower $\WHa^\mathrm{obs}$. The strongest predictor of large $L_\mathrm{IFS}/L_\mathrm{G}$ is the scatter at large $l^\mathrm{obs}_j/\langle l^\mathrm{obs}_j\rangle$ ($\sigma_u$, panel b), which has a Spearman rank correlation coefficient of 0.405. This is similar to what is seen in 8595-12702 above, and may be indicative of an increased DIG contribution driving the high values of $L_\mathrm{IFS}/L_\mathrm{G}$ in low sSFR galaxies. We see a strong correlation between the intrinsic width of the luminosity distribution, $\sigma$, and $\WHa^\mathrm{obs}$, but no correlation with $L_\mathrm{IFS}/L_\mathrm{G}$ (panel c). Finally, for a sanity check, in panel (d) we verify that the galaxies with the highest $L_\mathrm{IFS}/L_\mathrm{G}$ are not simply those with the lowest \Hb fluxes, and therefore are not likely to be caused by measurement errors. 

An additional complication with this analysis is that, in order to estimate $e^{+\tau_j}$ from the observed Balmer decrement, we must assume a dust attenuation law (we have used \citetalias{Cardelli.Clayton.Mathis.1989a}). If this is incorrect, we will have introduced a bias in  $e^{+\tau_j}$. Of more concern, if the slope of the attenuation law changes as a function of \Ha luminosity, then our measured values of $a$ will be incorrect. Let us consider a \citet{Wild.etal.2007a} dust law designed to correct emission lines which is proportional to $\lambda^{-1.3}$ for birth clouds (BCs) and to $\lambda^{-0.7}$ for the intervening interstellar medium (ISM) or the DIG. The different steepness of the dust attenuation law in the DIG can be partly attributed to more scattering of blue photons into the line of sight, and partly due to lower optical depth, meaning less blue light would be blocked by the ISM \citep[see e.g.][]{Chevallard.etal.2013a}. For the same $\tau$, we expect a smaller $\Ha/\Hb$ towards the DIG than towards BCs; if we then use a single average attenuation law we \emph{underestimate} $\tau$ in the DIG and \emph{overestimate} it in BCs. This would make the $l^\mathrm{obs}_j/\langle l^\mathrm{obs}_j\rangle$ versus $e^{+\tau_j} / \langle e^{+\tau_j} \rangle$ relation steeper than if we had used different laws for spaxels depending on their DIG contribution. The fact that we observe the largest positive values of $a$ for galaxies with smaller \WHa may indicate DIG at work. Looking back at Fig.~\ref{fig:toy-models}, moving $a$ closer to zero typically reduces $L_\mathrm{IFS}/L_\mathrm{G}$. Clearly effort is warranted to determine the slope of the attenuation curve in spatially-resolved observations, in order to accurately estimate the deficit in $L_\mathrm{G}$ measured from integrated observations. While this is not easy from emission line measurements alone, good quality observations of the stellar continuum could bring additional constraints.

\begin{figure}
  \centering
  \includegraphics[width=.8\columnwidth, trim=40 280 525 40, clip]{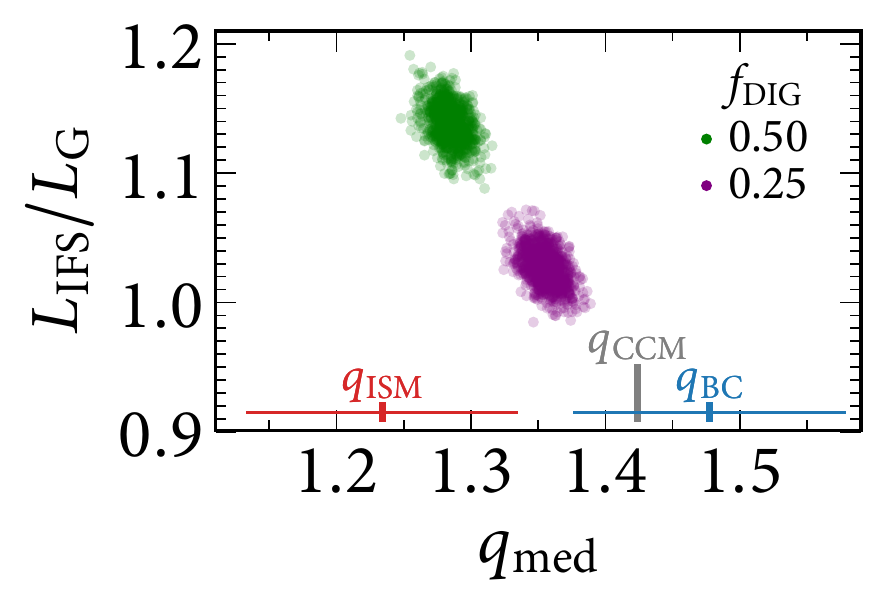}
  \caption{Toy models to investigate the effect of the DIG on $L_\mathrm{IFS}/L_\mathrm{G}$. We show two sets of models, each one with 1000 realisations, colour-coded by the fraction $f_\mathrm{DIG}$ of the observed \Ha line luminosity attributed to DIG emission. $L_\mathrm{IFS}/L_\mathrm{G}$ is shown as a function of the median 
steepness $q_j$ of the dust law for the spaxels within each realisation. At the bottom, we mark the value of $q$ for the \citetalias{Cardelli.Clayton.Mathis.1989a} law used to correct the global luminosities, and we show the centre and dispersion of the normal distributions used to randomly draw $q$ values for birth cloud (BC) and ISM spaxels.
    $L_\mathrm{IFS}/L_\mathrm{G}$ values are larger for the $f_\mathrm{DIG} = 0.50$ set of realisations, where the global $q$ is a poor descriptor of the individual values inside the galaxy.
  }
\label{fig:toyDIG}
\end{figure}

In order to further investigate the role of the DIG, we have created a set of toy models to gauge how much the point-by-point variation in the dust law slopes $q_j$ affects $L_\mathrm{IFS}/L_\mathrm{G}$ for objects with different fractional contributions from DIG ($f_\mathrm{DIG}$) to the observed \Ha line luminosity.
We create 1000 galaxy realisations with $f_\mathrm{DIG} = 0.50 \pm 0.01$ (mean and standard deviation)  and another set of 1000 realisations with $f_\mathrm{DIG} = 0.25 \pm 0.01$. Each galaxy realisation contains 200 spaxels. For each spaxel we randomly choose $e^{+\tau_j}$ from a Gaussian distribution with mean 2.0 and dispersion 0.1 (setting $1.0$ as the lower limit for $e^{+\tau_j}$), and its contribution to the total observed \Ha luminosity $w_j \equiv l_\mathrm{j}^\mathrm{obs} / \sum_j l_\mathrm{j}^\mathrm{obs}$ is drawn from a uniform distribution.
We draw $q_j$ values from a normal distribution with a dispersion of $0.1$; for the brightest spaxels in a given galaxy realisation, the centre of the distribution is $q_\mathrm{BC} = 1.48$ (corresponding to a $\lambda^{-1.3}$ BC dust law), while for the faintest ones it is $q_\mathrm{ISM} = 1.23$ (corresponding to a $\lambda^{-0.7}$ ISM dust law). The number of faint spaxels to be attenuated by an ISM dust law is chosen so that their attenuated luminosities match the input fraction $f_\mathrm{DIG}$ of the total observed luminosity; the remaining bright spaxels are assigned a BC dust law.
We set the $q = 1.42$ from the \citetalias{Cardelli.Clayton.Mathis.1989a} law to correct for dust in the global spectrum.

Fig.~\ref{fig:toyDIG} shows $L_\mathrm{IFS}/L_\mathrm{G}$ calculated from equation~(\ref{eq:LIFS_LG}) as a function of the median $q_j$ in each galaxy realisation. On the bottom of the plot we mark with a vertical grey line the value of $q = q_\mathrm{CCM}$ used to compute $L_\mathrm{G}$. The horizontal red and blue lines plus the thick line at the centre mark the dispersion and  mean  of the $q_\mathrm{BC}$ and $q_\mathrm{ISM}$ normal distributions from which the individual $q_j$s where drawn.
Realisations with $f_\mathrm{DIG} = 0.50$ (green dots) have median $q$ values closer to $q_\mathrm{ISM}$, while those with $f_\mathrm{DIG} = 0.25$ (purple dots) are closer to $q_\mathrm{BC}$.
There are two main lessons to take away from these toy models. The first is that, for $f_\mathrm{DIG} = 0.25$, $L_\mathrm{IFS}/L_\mathrm{G}$ values  fall in the $\sim 1.00$--$1.05$ range, which matches the range of values we have empirically found for MaNGA observations. Note that some points do fall below the $L_\mathrm{IFS}/L_\mathrm{G} < 1$ prediction, which may indicate that for the very few (3 out of 156) MaNGA galaxies in which the measurements of this ratio is smaller than one they may have a large $q_\mathrm{med}$ and small $f_\mathrm{DIG}$.
The second is that, for larger $f_\mathrm{DIG}$, $L_\mathrm{IFS}/L_\mathrm{G}$ is also larger because the $q$ value used for the global spectrum is no longer representative of the range of $q_j$s within the galaxy. This means larger values $L_\mathrm{IFS}/L_\mathrm{G}$ may be driven by the DIG contribution to emission line luminosities, which likely affects our low-$\WHa^\mathrm{obs}$ MaNGA galaxies.

In summary, we conclude that the DIG can cause a large $L_\mathrm{IFS}/L_\mathrm{G}$ due to two different effects. The first is by increasing the scatter in the dust attenuation--luminosity relation. This affects mostly galaxies with low sSFR, where spaxels with significant DIG emission contribute significantly to the total \Ha luminosity. The second effect is that a significant contribution from DIG may also induce an incorrect estimation of $L_\mathrm{IFS}$ by causing the dust attenuation law slope to vary with \Ha luminosity. Any sort of variation of slope with galaxy parameters has important wide-reaching implications, but is beyond the scope of this paper.

\subsection{WARPFIELD-POP simulation}
\label{sec:simulations}

\begin{figure*}
  \centering
  \includegraphics[width=0.8\textwidth, trim=180 40 270 70, clip]{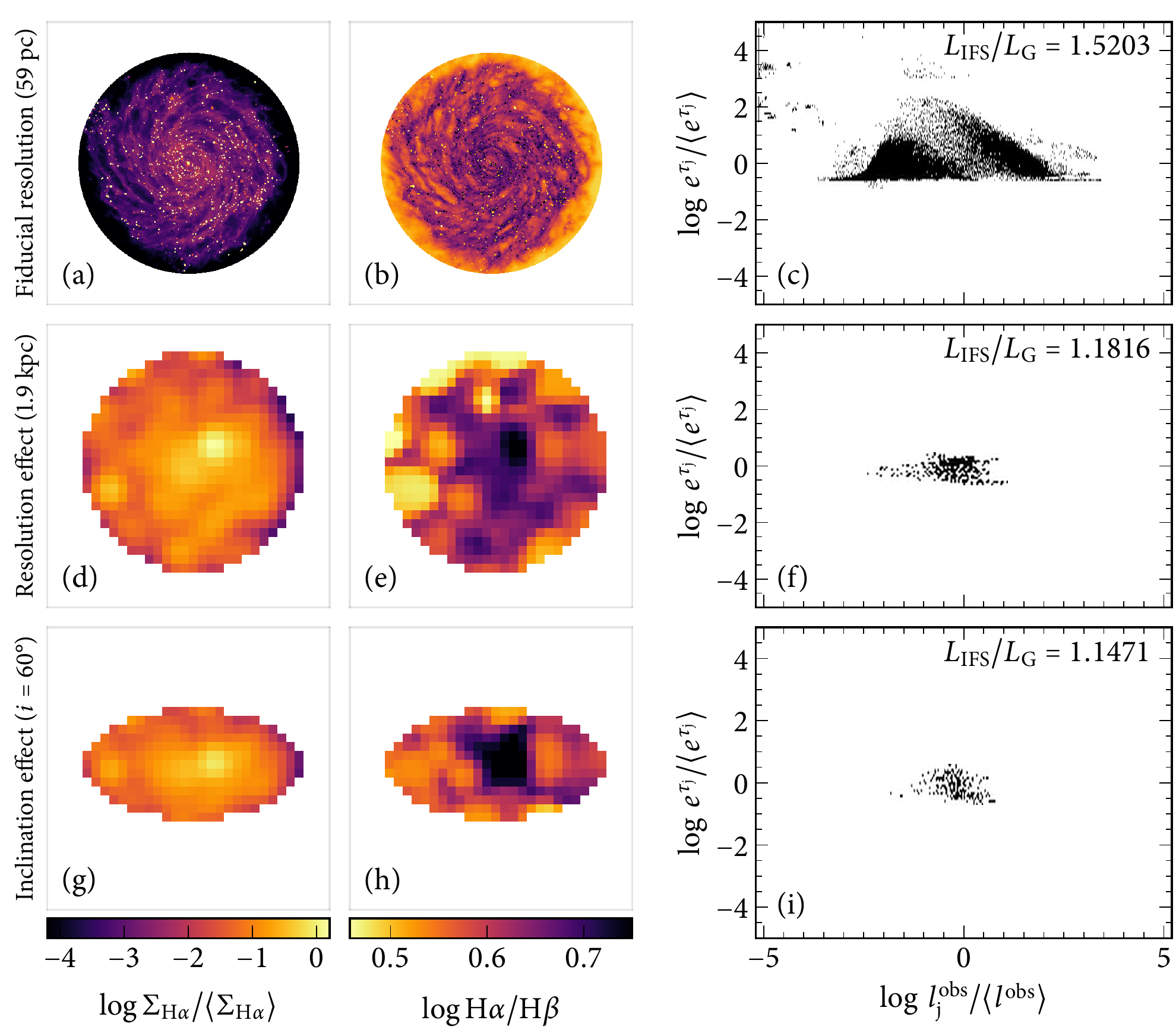}
  \caption{Top row: maps of intrinsic \Ha surface density and \Ha/\Hb for the face-on simulated {\scshape warpfield} galaxy with 59\,pc resolution, with the distribution of relative \Ha luminosity vs. dust attenuation for all spaxels on the right.  Centre row:  the same for the degraded image with 1.9\,kpc resolution. Bottom row: as for the centre row, but with the galaxy viewed at $i = 60^\circ$.
    The simulation has been limited to a diameter of 47~kpc. The $L_\mathrm{IFS}/L_\mathrm{G}$ ratio is given in the  top right of the scatter plots, and decreases from $\sim 52$ per cent for the original simulated galaxy to $\sim 18$ and $\sim 15$ per cent for the degraded face-on and edge-on galaxy.}
\label{fig:wf-maps}
\end{figure*}

Finally, we tackle the question of how $L_\mathrm{IFS}/L_\mathrm{G}$ might be expected to vary with inclination angle and observed spatial resolution by analysing a galaxy simulation post-processed with the {\scshape warpfield-pop} code \citep{Pellegrini.etal.2019a, Pellegrini.etal.2019c}. This has the advantage over our toy models of a semi-realistic spatial distribution of line luminosity and dust attenuation, although it does not allow us to probe changes in the line luminosity--dust attenuation relation, or in the shape of the attenuation curve. Galaxy Au-6 with a stellar mass of $6\times10^{10}$M$_\odot$ was selected from the Auriga cosmological magneto-hydrodynamical zoom-in simulations of Milky Way-like galaxies \citep{Grand.etal.2017a}. While the effective resolution of $\sim100$~pc in the highest density regions is very high by the standards of cosmological simulations, it is too low to follow the formation of stellar clusters or accurately model the impact of stellar feedback. The {\scshape warpfield-pop} post-processing code was therefore used to include spherically symmetric\footnote{A fair comparison to real data would require sophisticated models with 3D geometry, but there is none available at this resolution on full galaxy scales at the moment.} star-forming regions, with an age and mass distribution for star clusters consistent with the global star formation rate of the Milky Way. The cluster positions were randomly sampled, with the constraint that a cluster of mass $M$ could only be placed in a location where there is sufficient mass within 50\,pc to form a molecular cloud of mass $M_{\rm cl} = 100 M$.
Emission line emissivities are computed with {\scshape cloudy} v17 \citep{Ferland.etal.2017a}, stellar feedback is taken into account \citep{Rahner.etal.2017a, Rahner.etal.2018a}, and emission line photons are propagated through the ISM and attenuated by dust grains, assuming standard Milky Way grain distribution and composition, using the {\scshape polaris}\footnote{\url{http://www1.astrophysik.uni-kiel.de/~polaris/}} radiative transfer code \citep*{Reissl.Wolf.Brauer.2016a}. This final code additionally produces synthetic emission line and attenuation maps (see also \citealp{Pellegrini.etal.2019c}).  

Our analysis uses the intrinsic and dust-attenuated \Ha and \Hb emission line maps with 59\,pc pixels and a range of observed inclinations.  To create a simulated galaxy with MaNGA-like spatial resolution, we downgraded the maps by convolving them with a Gaussian point spread function with a standard deviation of 32 pixels (1.9\,kpc), then binned the cubes by $32 \times 32$ pixels. We retain only the central regions of the galaxy, 47\,kpc across, to account for surface brightness limitations in real data. Fig.~\ref{fig:wf-maps} shows original and downgraded maps of the intrinsic \Ha surface density and \Ha/\Hb for the face-on ($i = 0^\circ$) simulation, plus the downgraded maps for the most edge-on simulation ($i = 60^\circ$) still comparable to our MaNGA star-forming sample. In the right-hand panels we show the relation between relative \Ha luminosity and dust attenuation for all spaxels. As previously, we calculated $L_\mathrm{IFS}$ and $L_\mathrm{G}$ from the observed Balmer decrement using a \citetalias{Cardelli.Clayton.Mathis.1989a} dust law. 
The simulation uses a dust grain distribution to account for dust attenuation \citep{Reissl.Wolf.Brauer.2016a, Reissl.etal.2018a}, whereas our correction is limited by the use of an effective attenuation law. This is a limitation of all methods using a dust attenuation correction based on a single Balmer decrement ratio, as we can only hypothesize the shape of the appropriate effective attenuation law to be used. We note that the floor in $e^{+\tau_j} / \langle e^{+\tau_j} \rangle$ for the high resolution map is caused by $\tau_j$ reaching zero\footnote{In detail $e^{+\tau_j}$ reaches 1.03 due to a slight difference in the true dust law and/or intrinsic \Ha/\Hb line ratio and those assumed in our analysis.}.

\begin{figure}
  \centering
  \includegraphics[width=0.9\columnwidth, trim=60 350 560 20, clip]{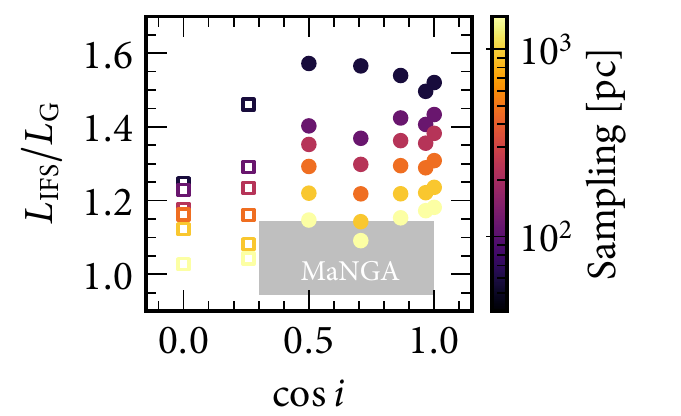}
  \caption{$L_\mathrm{IFS}/L_\mathrm{G}$ for the simulated {\scshape warpfield} galaxy viewed at different inclination angles ($i$). Different colours represent different spatial samplings (which are the set to vary with the spatial resolution), ranging from 59\,pc and doubling for each consecutive point up to 1.9\,kpc.  Very edge-on systems are plotted as open squares, as these would have been excluded from our MaNGA sample ($\cos i \sim b/a \ge 0.3$). The grey shaded area indicates the range of values in our MaNGA SF sample.
  }
\label{fig:wf}
\end{figure}

Fig.~\ref{fig:wf} shows $L_\mathrm{IFS}/L_\mathrm{G}$ for the original and spatially downgraded maps. We have created a set of downgraded maps using bin sizes of 2, 4, 8, 16 and 32 (corresponding to spatial resolutions of 0.12, 0.23, 0.47, 0.93 and 1.9~kpc respectively). Each version was downgraded by convolving the original maps with a Gaussian point spread function with a standard deviation of the bin size and then resampling the map.
For systems with $\cos i>0.5$ we find $L_\mathrm{IFS}/L_\mathrm{G} = 1.50$--$1.57$ for the original cubes and $1.09$--$1.18$ for the lowest spatial resolution cubes. These values are a little higher than we might expect compared to our MaNGA and MUSE results. We find little variation with inclination angle except for extreme edge-on systems where $L_\mathrm{IFS}$ becomes less informative and we approach $L_\mathrm{IFS}/L_\mathrm{G}=1$, which is probably due to \Hb (and possibly \Ha) being too attenuated and thus probing a smaller volume of the galaxy. Similar to our MUSE results, we see no evidence for saturation in the effect at the spatial scales probed, indicating that smaller and smaller observed scales might reveal a larger and larger discrepancy. 

Although we have focussed here on the observable effect of integrated versus spatially resolved \Ha luminosity, an obvious advantage of the simulations is the access to the intrinsic \Ha luminosity. We find that if we assume an effective dust attenuation curve with $R_V = 3.1$ and an intrinsic $\Ha/\Hb$ ratio of $2.87$ as done throughout this paper, $L_\mathrm{IFS}$ still underestimates the intrinsic luminosity by a factor of $1.5$--$2.1$ at the native resolution of the simulation of 59\,pc (the underestimation is even greater for inclinations of $\cos i < 0.5$). Further investigation shows that both the intrinsic $\Ha/\Hb$ ratio and effective attenuation curve shape changes with column density in the simulation, with higher density regions (i.e.\ those in the spiral arms or towards the centre of the galaxy) having flatter attenuation curve shapes (i.e. $R_V > 3.1$) and more diffuse regions having higher intrinsic $\Ha/\Hb$ (with values up to $\sim 3.2$). The most significant effect is caused by the very flat attenuation curves in the dense star-forming regions, leading to $R_V$ values that are at odds with observational results, albeit these are averaged over large areas \citep[e.g.][]{Wild.etal.2011a}.  It is possible that with high enough spatial resolution observations, such as those with SITELLE \citep{Brousseau.etal.2014a, Drissen.etal.2019a} or SDSS-V Local Volume Mapper \citep{Kollmeier.etal.2017a}, this effect could be measured directly.

\section{Summary}
\label{sec:summary}

From a sample of over 4800 MaNGA observations, and after carefully removing their stellar continuum via spectral fitting, we select 156 star-forming galaxies with high quality \Ha and \Hb emission line detections. For each galaxy, we compare their dust-corrected \Ha luminosity obtained from the global integrated spectrum ($L_\mathrm{G}$) to the sum of \Ha luminosities dust-corrected spaxel-by-spaxel ($L_\mathrm{IFS}$). We perform the same analysis on 12 non-overlapping MUSE datacubes of NGC\,628, both at the native spatial sampling of 36\,pc and downgraded down to 0.58--1.2\,kpc, similar to the MaNGA observations. 

As inferred by simple analytical arguments, we find $L_\mathrm{IFS}/L_\mathrm{G} \ge 1$ within the errors for 153/156 MaNGA galaxies, with a difference of 2--4 per cent on average, and as much as 15 per cent for some galaxies. Given the limited spatial resolution of MaNGA, we note that this is only a lower limit on the true effect, as each spaxel itself is an integration over many star-forming regions.
Also, our MaNGA sample comprises high-quality normal SF galaxies, so the discrepancy for lower-sSFR galaxies and LIRGs could be much higher.
Measuring $L_\mathrm{IFS}/L_\mathrm{G}$ on NGC\,628 at a resolution of 36\,pc and also degraded to the MaNGA-like resolution, we observe a variation from 14 to 0--3 per cent respectively. We find no saturation at high spatial resolutions, indicating the true effect may be even larger. This result demonstrates the importance of spatial sampling in (i) deriving the true intrinsic line luminosities, and (ii) quantifying a point-to-point versus global correction factor, which is crucial to correct the underestimation of intrinsic luminosities in studies that do not benefit from high spatial resolution, for example at higher-redshift.

Both a simple analytical approach and toy models demonstrate that $L_\mathrm{IFS}/L_\mathrm{G} \ge 1$ for all combinations of dust attenuation law slope and dust attenuation--line luminosity relation. However, we have found no single observable that determines what value $L_\mathrm{IFS}/L_\mathrm{G}$ will take, making it difficult to predict it for an individual galaxy. We show that both increasing the slope of the dust attenuation--line luminosity relation, as well as increasing the scatter in dust attenuation at large line luminosities, may lead to large values of $L_\mathrm{IFS}/L_\mathrm{G}$. All other parameters, such as the attenuation law slope, scatter in dust attenuation at lower line luminosities and width of the line luminosity distribution become more important if the slope of the relation is large. 

We found that galaxies in the MaNGA sample showing the largest values of $L_\mathrm{IFS}/L_\mathrm{G}$ have low  $\WHa^\mathrm{obs}$. In one example case, we showed that this was due to an off-centre relatively dust-free star-forming region that was contributing significantly to the total luminosity of the galaxy. The increased scatter in $L_\mathrm{IFS}/L_\mathrm{G}$ may be caused by low sSFR galaxies being more likely to have such unusual geometries than high sSFR galaxies. However, we also show that an increased scatter in the dust attenuation-line luminosity relation could arise from significant contribution from the diffuse ionised gas in low sSFR galaxies, and that the contribution of the DIG may also affect the dust attenuation law slope.

To complete our study, we performed the same analysis on a simulated galaxy, with emission lines generated with {\scshape warpfield-pop}, demonstrating how $L_\mathrm{IFS}/L_\mathrm{G}$ varies smoothly with spatial scale, with little variation with inclination for mostly face-on galaxies ($\cos i>0.5$). We find that the ratio approaches unity at high inclinations, but this is because $L_\mathrm{IFS}$ becomes an increasingly poor measurement of the total \Ha luminosity as might be expected. As with the MUSE observations of NGC\,628 there is no evidence for a saturation in the ratio at the smallest scales probed by the simulation, which could suggest that the true discrepancy might be $>$50 per cent, although we caution that the ratios measured in the simulation are noticeably larger than those observed in both our MaNGA and MUSE galaxies. Observations such as those presented in this paper should provide new constraints on simulations in the future. 

As the study of galaxy evolution demands increasingly accurate measurements of the physical properties of galaxies, our results have important implications for estimating accurate star formation rates for galaxies in the local and distant Universe, as well as gas-phase metallicities from certain line combinations.
Our study has revealed the complexity of accurate \Ha measurements and enabled us to quantify the variations of $L_\mathrm{IFS}/L_\mathrm{G}$ (1) within a large galaxy sample, (2) with spatial resolution. We have investigated a series of parameters with which the ratio could correlate, and found that the star formation history and the DIG contribution are important.

Upcoming surveys such as the SDSS-V Local Volume Mapper \citep{Kollmeier.etal.2017a} and SIGNALS \citep{Rousseau-Nepton.etal.2019a} will provide very high spatial resolution integral field spectroscopic observations of many nearby galaxies. These surveys will undoubtedly provide further important insights into this problem, potentially tackling aspects that we are unable to reach with current data, such as comparison to true \Ha luminosities and internal variations of the attenuation curve slope.

\section*{Acknowledgements}  

    We thank Roberto Cid Fernandes, Gra\.zyna Stasi\'nska and Kenny Wood for fruitful discussions, Kathryn Kreckel for the help aligning the MUSE datacubes, Fabio Bresolin and Danielle Berg for the details on \hii region observations used on an earlier draft, and Dmitry Bizyaev, M\'ed\'eric Boquien and Gra\.zyna Stasi\'nska for useful comments on an earlier version of this manuscript.

    NVA would like to thank the University of St Andrews for providing support during her visit. NVA acknowledges support of the Royal Society and the Newton Fund via the award of a Royal Society--Newton Advanced Fellowship (grant NAF\textbackslash{}R1\textbackslash{}180403), and of Funda\c{c}\~ao de Amparo \`a Pesquisa e Inova\c{c}\~ao de Santa Catarina (FAPESC) and Conselho Nacional de Desenvolvimento Cient\'{i}fico e Tecnol\'{o}gico (CNPq).
    AW acknowledges financial support from Funda\c{c}\~ao de Amparo \`a Pesquisa do Estado de S\~ao Paulo (FAPESP) process number 2019/01768-6.
    MG receives funding from the European Research Council (ERC) under the European Union Horizon 2020 research and innovative programme (MagneticYSOs programme, grant agreement Nber 679937).
    EWP, RSK, SR, SCOG and DR acknowledge funding from the Deutsche Forschungsgemeinschaft (DFG) via the Collaborative Research Center (SFB 881) `The Milky Way System' (subprojects A1, B1, and B2) and from the Heidelberg Cluster of Excellence {\em STRUCTURES} in the framework of Germany's Excellence Strategy (grant EXC-2181/1 - 390900948).

    Funding for the Sloan Digital Sky Survey IV has been provided by the Alfred P. Sloan Foundation, the U.S. Department of Energy Office of Science, and the Participating Institutions. SDSS-IV acknowledges support and resources from the Center for High-Performance Computing at the University of Utah. The SDSS web site is www.sdss.org.
    SDSS-IV is managed by the Astrophysical Research Consortium for the Participating Institutions of the SDSS Collaboration including the Brazilian Participation Group, the Carnegie Institution for Science, Carnegie Mellon University, the Chilean Participation Group, the French Participation Group, Harvard-Smithsonian Center for Astrophysics, Instituto de Astrof\'isica de Canarias, The Johns Hopkins University, Kavli Institute for the Physics and Mathematics of the Universe (IPMU) / University of Tokyo, the Korean Participation Group, Lawrence Berkeley National Laboratory, Leibniz Institut f\"ur Astrophysik Potsdam (AIP), Max-Planck-Institut f\"ur Astronomie (MPIA Heidelberg), Max-Planck-Institut f\"ur Astrophysik (MPA Garching), Max-Planck-Institut f\"ur Extraterrestrische Physik (MPE), National Astronomical Observatories of China, New Mexico State University, New York University, University of Notre Dame, Observat\'ario Nacional / MCTI, The Ohio State University, Pennsylvania State University, Shanghai Astronomical Observatory, United Kingdom Participation Group, Universidad Nacional Aut\'onoma de M\'exico, University of Arizona, University of Colorado Boulder, University of Oxford, University of Portsmouth, University of Utah, University of Virginia, University of Washington, University of Wisconsin, Vanderbilt University, and Yale University.
    
    This research made use of Astropy,\footnote{Astropy {\scshape python} package: \url{http://www.astropy.org}.} a community-developed core Python package for Astronomy \citep{AstropyCollaboration.etal.2013a, AstropyCollaboration.etal.2018a}.

\section{Data availability}

The data concerning MaNGA global results (i.e.\ those required to be able to reproduce Figs.~\ref{fig:LHa-corr}, \ref{fig:AD} and \ref{fig:AD-lq} in this article) are available in Zenodo at \url{https://dx.doi.org/10.5281/zenodo.3981027}.
This dataset was derived from the publicly available 15th SDSS data release (\citealp{Aguado.etal.2019a}; \url{https://www.sdss.org/dr15/data_access}).


\bibliography{references}

\appendix
\section[]{$L_\mathrm{IFS}/L_\mathrm{G}$ trends with other physical and observational parameters}
\label{app:stats}

Our requirement of high-quality data for each spaxel biases our sample against low-\WHa MaNGA galaxies. This appendix shows which objects are underrepresented in our study, and investigates whether other galaxy properties are correlated to $L_\mathrm{IFS}/L_\mathrm{G}$ trends.

\begin{figure}
  \centering
  \includegraphics[width=1\columnwidth, trim=40 220 400 40]{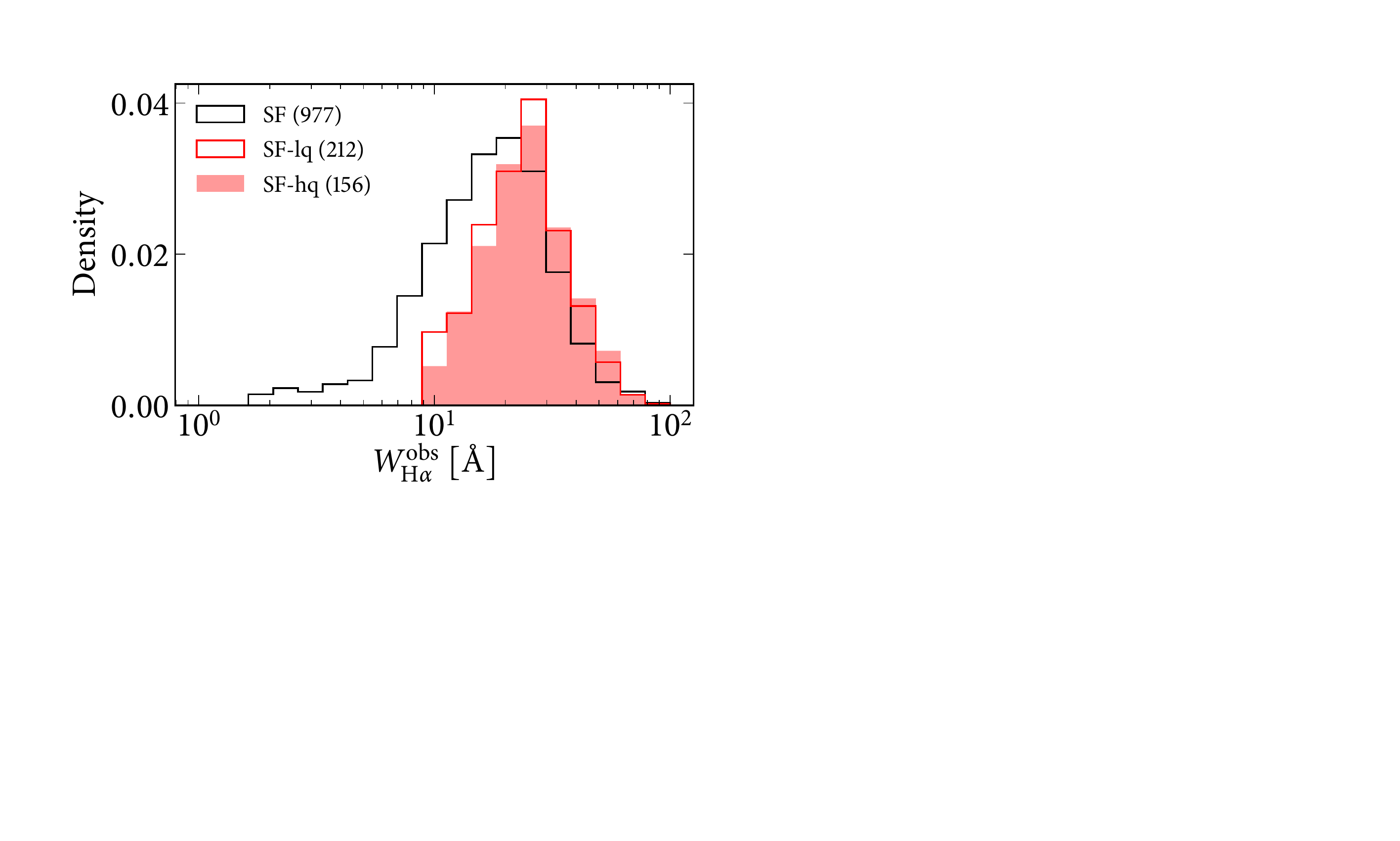}
  \caption{Distribution of the integrated \Ha equivalent width for the full (SF, black outline), low-quality (SF-lq, red outline) and high-quality (SF-hq, red filled region) samples of MaNGA star-forming galaxies. The number of objects in each sample is given in brackets.}
\label{fig:WHa}
\end{figure}

Let us first quantify how representative our MaNGA sample is when compared to less restrictive selections.
We refer to Section~\ref{sec:sample} for criteria used to define our master sample of 3\,185 galaxies.
Our SF sample is constituted by the 977 objects for which emission lines ratios measured in the integrated spectra lie in the pure star-forming region delimited by the \cite{Stasinska.etal.2006a} line in the \Nii/\Ha versus \Oiii/\Hb plane.
Out of those, we define the high-quality (SF-hq) sample studied in this paper, containing 156 star-forming galaxies where $A/N > 2$ in \Ha and \Hb for all spaxels.
For comparison, we define a low-quality (SF-lq) sample where $A/N > 1$ in \Ha and \Hb for all spaxels. Removing one object for which the $L_\mathrm{IFS}/L_\mathrm{G}$ uncertainty is $> 1$, the SF-lq sample comprises 212 galaxies.
Fig.~\ref{fig:WHa} shows the distribution of the integrated observed \WHa for the SF, SF-lq and SF-hq samples.
The SF-hq sample misses $\WHa < 20$~\AA\ objects when compared to the SF sample. The SF-lq sample is still biased against low-\WHa objects, but contains a few more objects where $\WHa < 30$~\AA.

\begin{figure*}
  \centering
  \includegraphics[width=\textwidth, trim=60 180 40 10]{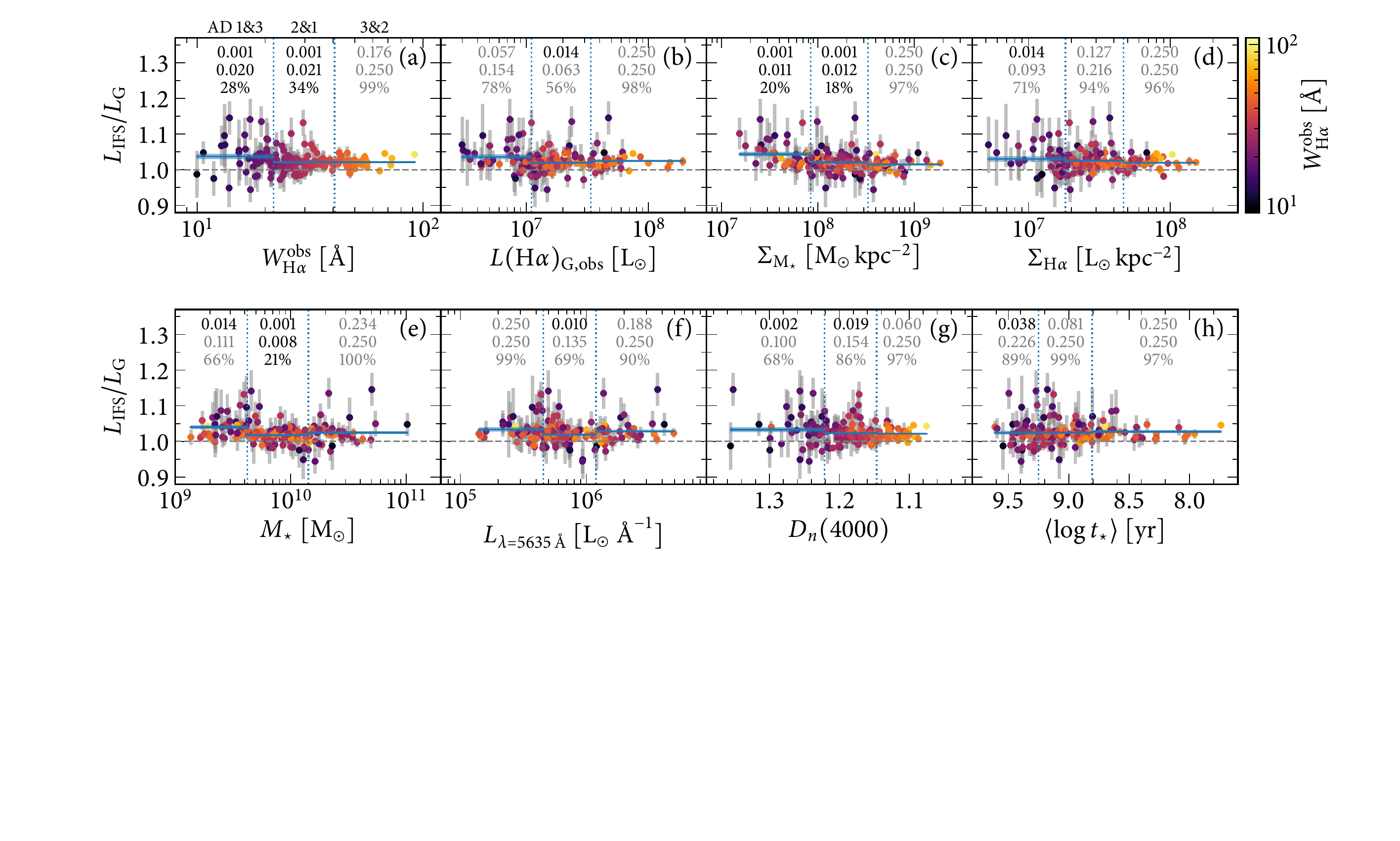}
  \caption{$L_\mathrm{IFS}/L_\mathrm{G}$ as a function of galaxy properties measured in the integrated spectra and colour-coded by \WHa for the SF-hq sample. (a) \WHa (similar to Fig.~\ref{fig:LHa-corr}),
(b) observed \Ha luminosity,
(c) stellar mass surface density,
(d) observed \Ha surface density,
(e) stellar mass, 
(f) observed luminosity at $\lambda = 5636$~\AA\,
(g) 4000-\AA\ break strength 
and (h) luminosity-weighted stellar population age.
Panels (g) and (h) have their $x$-axes drawn backwards so that objects on the left- and right-hand side roughly correspond to low and high \WHa values.
The horizontal dashed line marks $L_\mathrm{IFS}/L_\mathrm{G} = 1$, and the dotted vertical lines delimit the three bins in each panel (bin 1 is the first quartile, bin 2 the two central quartiles, and bin 3 the last quartile).
Numbers on top of each panel refer to $p$-values from the Anderson-Darling test that bins 1\&3, 2\&1 and 3\&2 have been drawn from the same distribution. The first line shows results for the test run on the original data, the second line shows the median $p$ value out of 100 realisations for perturbed data, and the third line shows the percentage of realisations where $p > 0.05$.
To highlight small $p$ values (i.e.\ low probability that the two bins are drawn from the same distribution), numbers are in grey when $p > 0.05$ or the fraction is $> 50$ per cent.
Note that $p$ values are capped at $0.001$ and $0.250$.
}
\label{fig:AD}
\end{figure*}

Given that our SF-hq sample misses low-\WHa objects, we revisit our results comparing $L_\mathrm{IFS}/L_\mathrm{G}$ values for the lowest and highest \WHa quartiles, which were based on Fig.~\ref{fig:LHa-corr}.
Fig.~\ref{fig:AD} shows $L_\mathrm{IFS}/L_\mathrm{G}$ as a function of a few observational and physical properties, all colour-coded by \WHa. The horizontal dashed line marks $L_\mathrm{IFS}/L_\mathrm{G} = 1$. All panels show properties which have been measured in the integrated spectra:
(a) $\WHa^\mathrm{obs}$, i.e. similar to Fig.~\ref{fig:LHa-corr},
(b) observed \Ha luminosity,
(c) stellar mass surface density (where the stellar mass has been measured from the optical spectral fitting using the \starlight code as in \citealp{CidFernandes.etal.2005a}),
(d) observed \Ha surface density,
(e) stellar mass, 
(f) observed luminosity at $\lambda = 5635 \pm 45$~\AA\ (a feature-free spectral region near the centre of the $V$-band),
(g) 4000-\AA\ break strength (larger values indicate older stellar populations)
and (h) luminosity-weighted stellar population age measured from the spectral synthesis \citep[as in][]{CidFernandes.etal.2005a}.
The $x$-axes in panels (g) and (h) are drawn backwards so that objects on the left- and right-hand side roughly correspond to low and high \WHa values.

For each integrated property, the SF-hq sample has been divided into quartiles, from which we define three bins: (1) the lowest quartile, (2) the two central quartiles, and (3) the highest quartile.
Vertical dotted lines show bin limits, whereas horizontal solid lines and filled regions show the average $L_\mathrm{IFS}/L_\mathrm{G}$ and its uncertainty within a bin.
Table~\ref{tab:stats-hq} shows, for each parameter in Fig.~\ref{fig:AD}, the limit between the three bins, and also the average $L_\mathrm{IFS}/L_\mathrm{G}$ value and its uncertainty $\Delta$ for each bin.
The uncertainty concerns only the propagation of errors in individual points, i.e. if the uncertainty in $L_\mathrm{IFS}/L_\mathrm{G}$ for each $k$ object is $\Delta y_k$, the uncertainty in $\langle L_\mathrm{IFS}/L_\mathrm{G} \rangle$ for a bin containing $n$ objects is $\Delta = \sqrt{\sum_{k=1}^{n} \Delta y_k^2} / n$.

We compare $L_\mathrm{IFS}/L_\mathrm{G}$ distributions among bins using the Anderson-Darling \citep[AD;][]{Press.etal.2007a} statistical test. The AD test gives the probability $p$ that two bins are drawn from the same distribution.
The results for the AD tests are at the top of each panel of Fig.~\ref{fig:AD}.
The first line shows $p$ values for AD tests between bins 1 and 3; 2 and 1; and 3 and 2. Numbers are shown in grey when $p > 0.05$ (two-sigma threshold) and in black otherwise, so as to highlight in a darker colour when bins are unlikely to come from the same distribution.
Note that the we use the \textsc{scipy} \citep{Virtanen.etal.2020a} implementation of this test (\texttt{scipy.stats.anderson\_ksamp}), which caps $p$ values below $0.001$ and above $0.250$.
In order to account for the effect of larger error bars at the left-hand side of plots (very notable at least for \WHa), we reshuffle $\Delta y_k$ uncertainties within the whole SF-hq sample and perturb the $L_\mathrm{IFS}/L_\mathrm{G}$ values accordingly.
We repeat this 100 times for each panel.
The second line of numbers thus shows the median $p$ (also in grey when $> 0.05$); and the third line is the percentage of realisations where $p > 0.05$ (in grey where $> 50$ per cent, which also implies a median $p > 0.05$).
Using different random seeds may change the median $p$ and percentages, but after auditing results for a handful of seeds we found that low median $p$ values remain consistently low from one set of realisations to the other.

Low $p$ values (i.e.\ low probability that bins come from the same distribution) can be found comparing bins 1 and 2, and 1 and 3 only for \WHa and the stellar mass density (panels a and c). 
Marginal results appear for all other quantities except for the observed luminosity at $5636$~\AA\ (panel f).
There seems to be no difference between bins 2 and 3 for any parameter.

\begin{figure*}
  \centering
  \includegraphics[width=\textwidth, trim=60 180 40 10]{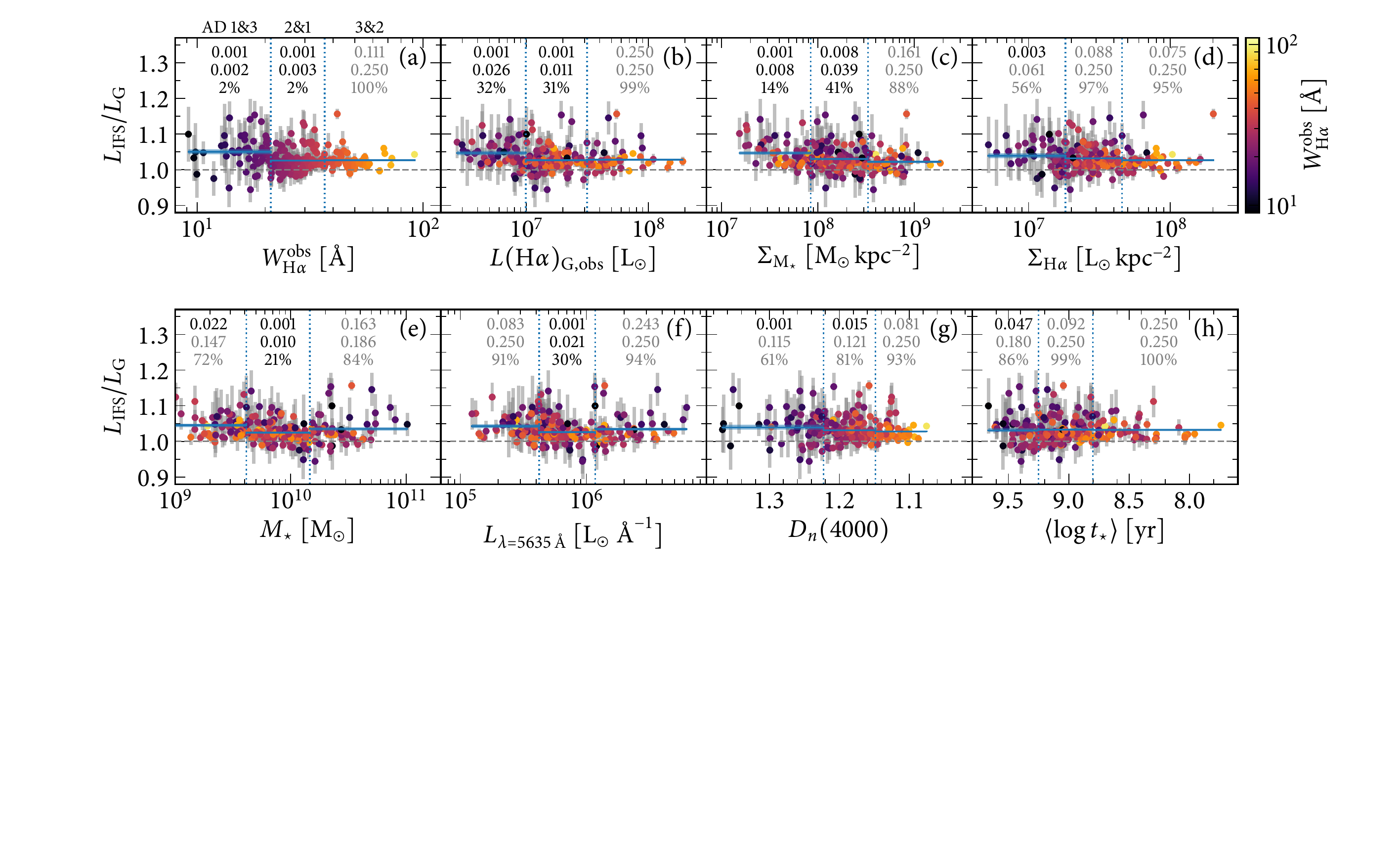}
  \caption{As Fig.~\ref{fig:AD}, but for the 212 objects in the SF-lq sample.}
\label{fig:AD-lq}
\end{figure*}

Fig.~\ref{fig:AD-lq} and Table~\ref{tab:stats-lq} show the same as Fig.~\ref{fig:AD} and Table Table~\ref{tab:stats-hq} but for the SF-lq sample.
Average $L_\mathrm{IFS}/L_\mathrm{G}$ values are larger at all \WHa bins, and the effect is even larger at the lowest \WHa bin.
Low AD $p$ values between bins 1 and 2, and 1 and 3 are found for \WHa, observed \Ha luminosity and stellar mass density (panels a, b and c). 
Marginal results appear again for all other parameters but for the observed luminosity at $5635$~\si{\angstrom}.
Inasmuch as properties are related to the specific SFR or DIG content, there seems to be a trend of larger $L_\mathrm{IFS}/L_\mathrm{G}$ values, and usually larger scatter in those values. 
We warn the reader however not to take the $L_\mathrm{IFS}/L_\mathrm{G}$ results from this study at face-value to correct their integrated \Ha luminosities. 
One must remember that (1) the MaNGA data has a coarse spatial resolution ($\sim 1$~kpc), so trends in $L_\mathrm{IFS}/L_\mathrm{G}$ shown here should be taken as lower limits of real values, and (2) deeper data would be needed to confirm trends for low-\WHa objects, which are underrepresented in our sample.

\begin{table*}
    \centering
    \caption{Statistics for Fig.~\ref{fig:AD} (SF-hq sample).}
    \begin{tabular}{lccccc}
        \hline
        (Panel) Parameter & Edge between bins 1\&2 & Edge between bins 2\&3 & Bin 1: $\langle L_\mathrm{IFS}/L_\mathrm{G} \rangle \pm \Delta$ & Bin 2: $\langle L_\mathrm{IFS}/L_\mathrm{G} \rangle \pm \Delta$ & Bin 3: $\langle L_\mathrm{IFS}/L_\mathrm{G} \rangle \pm \Delta$ \\
        \hline
        (a) $\WHa^\mathrm{obs}$ [\si{\angstrom}]                                            & $         21.8$ & $         40.6$ & $1.037 \pm 0.008$ & $1.021 \pm 0.003$ & $1.021 \pm 0.002$ \\
        (b) $L(\mathrm{H\alpha})_\mathrm{G,obs} \;\mathrm{[L_\odot]}$                       & $\num{1.10e+07}$ & $\num{3.38e+07}$ & $1.036 \pm 0.007$ & $1.020 \pm 0.003$ & $1.024 \pm 0.003$ \\
        (c) $\Sigma_\mathrm{M_\star} \;\mathrm{[M_\odot\, kpc^{-2}]}$                         & $\num{8.46e+07}$ & $\num{3.31e+08}$ & $1.043 \pm 0.006$ & $1.021 \pm 0.003$ & $1.015 \pm 0.003$ \\
        (d) $\Sigma_\mathrm{H\alpha} \;\mathrm{[L_\odot\, kpc^{-2}]}$                        & $\num{1.82e+07}$ & $\num{4.67e+07}$ & $1.030 \pm 0.007$ & $1.025 \pm 0.003$ & $1.019 \pm 0.003$ \\
        (e) $M_\mathrm{\star} \;\mathrm{[M_\odot]}$                                         & $\num{4.21e+09}$ & $\num{1.43e+10}$ & $1.040 \pm 0.006$ & $1.018 \pm 0.003$ & $1.024 \pm 0.004$ \\
        (f) $L_\mathrm{\lambda=5635\,\normalsize\AA} \;\mathrm{[L_\odot \;\si{\angstrom}^{-1}]}$       & $\num{4.56e+05}$ & $\num{1.19e+06}$ & $1.034 \pm 0.006$ & $1.019 \pm 0.003$ & $1.028 \pm 0.004$ \\
        (g) $D_n(4000)$                                                                  & $          1.22$ & $          1.15$ & $1.033 \pm 0.007$ & $1.023 \pm 0.003$ & $1.021 \pm 0.003$ \\
        (h) $\langle \log t_\star \rangle \;\mathrm{[yr]}$                               & $          9.25$ & $          8.81$ & $1.024 \pm 0.006$ & $1.025 \pm 0.003$ & $1.027 \pm 0.004$ \\
      \hline
    \end{tabular}
    \label{tab:stats-hq}
\end{table*}
 
\begin{table*}
    \centering
    \caption{Statistics for Fig.~\ref{fig:AD-lq} (SF-lq sample).}
    \begin{tabular}{lccccc}
        \hline
        (Panel) Parameter & Edge between bins 1\&2 & Edge between bins 2\&3 & Bin 1: $\langle L_\mathrm{IFS}/L_\mathrm{G} \rangle \pm \Delta$ & Bin 2: $\langle L_\mathrm{IFS}/L_\mathrm{G} \rangle \pm \Delta$ & Bin 3: $\langle L_\mathrm{IFS}/L_\mathrm{G} \rangle \pm \Delta$ \\
        \hline
        (a) $W_\mathrm{H\alpha}^\mathrm{obs} \;\mathrm{[\si{\angstrom}]}$                               & $         21.2$ & $         36.7$ & $1.051 \pm 0.007$ & $1.026 \pm 0.003$ & $1.027 \pm 0.002$ \\
        (b) $L(\mathrm{H\alpha})_\mathrm{G,obs} \;\mathrm{[L_\odot]}$                        & $\num{9.92e+06}$ & $\num{3.15e+07}$ & $1.047 \pm 0.006$ & $1.027 \pm 0.003$ & $1.028 \pm 0.003$ \\
        (c) $\Sigma_\mathrm{M_\star} \;\mathrm{[M_\odot\, kpc^{-2}]}$                        & $\num{8.46e+07}$ & $\num{3.31e+08}$ & $1.047 \pm 0.005$ & $1.030 \pm 0.003$ & $1.022 \pm 0.003$ \\
        (d) $\Sigma_\mathrm{H\alpha} \;\mathrm{[L_\odot\, kpc^{-2}]}$                        & $\num{1.82e+07}$ & $\num{4.58e+07}$ & $1.039 \pm 0.007$ & $1.032 \pm 0.003$ & $1.027 \pm 0.003$ \\
        (e) $M_\mathrm{\star} \;\mathrm{[M_\odot]}$                                          & $\num{4.14e+09}$ & $\num{1.46e+10}$ & $1.045 \pm 0.005$ & $1.024 \pm 0.003$ & $1.035 \pm 0.004$ \\
        (f) $L_\mathrm{\lambda=5635\,\normalsize\si{\angstrom}} \;\mathrm{[L_\odot \;\si{\angstrom}^{-1}]}$        & $\num{4.21e+05}$ & $\num{1.17e+06}$ & $1.043 \pm 0.005$ & $1.026 \pm 0.003$ & $1.034 \pm 0.004$ \\
        (g) $D_n(4000)$                                                                      & $          1.22$ & $          1.15$ & $1.039 \pm 0.007$ & $1.031 \pm 0.003$ & $1.028 \pm 0.003$ \\
        (h) $\langle \log t_\star \rangle \;\mathrm{[yr]}$                                   & $          9.25$ & $          8.80$ & $1.031 \pm 0.006$ & $1.033 \pm 0.003$ & $1.032 \pm 0.003$ \\      \hline
    \end{tabular}
    \label{tab:stats-lq}
\end{table*}

\bsp	

\label{lastpage}

\end{document}